\documentclass[12pt,english]{article}
\usepackage{avant}
\usepackage[T1]{fontenc}
\usepackage[latin9]{inputenc}
\usepackage{geometry}
\usepackage{amsmath}
\usepackage{amssymb}
\usepackage{stmaryrd}
\usepackage{babel}
\usepackage{array}
\usepackage{float}
\usepackage{multirow}
\usepackage{amsmath}
\usepackage{amssymb}
\usepackage{stmaryrd}
\usepackage[authoryear]{natbib}
\usepackage[unicode=true,
bookmarks=false,
breaklinks=false,pdfborder={0 0 1},colorlinks=false]
{hyperref}
\usepackage{breakurl}

\makeatletter

\floatstyle{ruled}
\newfloat{algorithm}{tbp}{loa}
\providecommand{\algorithmname}{Algorithm}
\floatname{algorithm}{\protect\algorithmname}

\makeatletter
\def\widebreve{\mathpalette\wide@breve}
\def\wide@breve#1#2{\sbox\z@{$#1#2$}%
     \mathop{\vbox{\m@th\ialign{##\crcr
\kern0.08em\brevefill#1{0.8\wd\z@}\crcr\noalign{\nointerlineskip}%
                    $\hss#1#2\hss$\crcr}}}\limits}
\def\brevefill#1#2{$\m@th\sbox\tw@{$#1($}%
  \hss\resizebox{#2}{\wd\tw@}{\rotatebox[origin=c]{90}{\upshape(}}\hss$}
\makeatletter

\usepackage{amsthm}\usepackage{bm}\usepackage{graphicx}\usepackage{rotating}\usepackage{array}\usepackage{booktabs}\usepackage{color}\usepackage{algorithmic}
\usepackage{algorithm}\usepackage{multirow}\usepackage{enumerate}\usepackage{graphicx}\usepackage{caption}\usepackage{subcaption}

\floatstyle{ruled}
\newfloat{algorithm}{tbp}{loa}
\providecommand{\algorithmname}{Algorithm}
\floatname{algorithm}{\protect\algorithmname}

\topmargin by -\headheight \advance
\topmargin by -\headsep

\evensidemargin
\oddsidemargin
\@ifundefined{definecolor}{color}{}
\RequirePackage[displaymath]{lineno}

\@ifundefined{definecolor}{color}{}
\usepackage{amsfonts}\usepackage{latexsym}

\usepackage{mathrsfs}\usepackage{amsthm}\usepackage{latexsym}\usepackage{booktabs}\@ifundefined{definecolor}{color}{}

\usepackage{times}\@ifundefined{definecolor}{color}{}

\numberwithin{equation}{section}
\numberwithin{figure}{section}

\newcommand{\T}{^{\mbox{\tiny {\sf T}}}}
\DeclareMathOperator{\diag}{diag}

\usepackage{babel}

\DeclareMathOperator*{\argmin}{argmin}

\makeatother

\begin{document}
	{\setlength{\baselineskip}{1.5\baselineskip}

		\global\long\def\mba{\mathbf{a}}
		\global\long\def\mbA{\mathbf{A}}
		\global\long\def\mbb{\mathbf{b}}
		\global\long\def\mbB{\mathbf{B}}
		\global\long\def\mbc{\mathbf{c}}
		\global\long\def\mbC{\mathbf{C}}
		\global\long\def\mbd{\mathbf{d}}
		\global\long\def\mbD{\mathbf{D}}
		\global\long\def\mbe{\mathbf{e}}
		\global\long\def\mbE{\mathbf{E}}
		\global\long\def\mbf{\mathbf{f}}
		\global\long\def\mbF{\mathbf{F}}
		\global\long\def\mbg{\mathbf{g}}
		\global\long\def\mbG{\mathbf{G}}
		\global\long\def\mbh{\mathbf{h}}
		\global\long\def\mbH{\mathbf{H}}
		\global\long\def\mbi{\mathbf{i}}
		\global\long\def\mbI{\mathbf{I}}
		\global\long\def\mbj{\mathbf{j}}
		\global\long\def\mbJ{\mathbf{J}}
		\global\long\def\mbk{\mathbf{k}}
		\global\long\def\mbK{\mathbf{K}}
		\global\long\def\mbl{\mathbf{l}}
		\global\long\def\mbL{\mathbf{L}}
		\global\long\def\mbm{\mathbf{m}}
		\global\long\def\mbM{\mathbf{M}}
		\global\long\def\mbn{\mathbf{n}}
		\global\long\def\mbN{\mathbf{N}}
		\global\long\def\mbo{\mathbf{o}}
		\global\long\def\mbO{\mathbf{O}}
		\global\long\def\mbp{\mathbf{p}}
		\global\long\def\mbP{\mathbf{P}}
		\global\long\def\mbq{\mathbf{q}}
		\global\long\def\mbQ{\mathbf{Q}}
		\global\long\def\mbr{\mathbf{r}}
		\global\long\def\mbR{\mathbf{R}}
		\global\long\def\mbs{\mathbf{s}}
		\global\long\def\mbS{\mathbf{S}}
		\global\long\def\mbt{\mathbf{t}}
		\global\long\def\mbT{\mathbf{T}}
		\global\long\def\mbu{\mathbf{u}}
		\global\long\def\mbU{\mathbf{U}}
		\global\long\def\mbv{\mathbf{v}}
		\global\long\def\mbV{\mathbf{V}}
		\global\long\def\mbw{\mathbf{w}}
		\global\long\def\mbW{\mathbf{W}}
		\global\long\def\mbx{\mathbf{x}}
		\global\long\def\mbX{\mathbf{X}}
		\global\long\def\mby{\mathbf{y}}
		\global\long\def\mbY{\mathbf{Y}}
		\global\long\def\mbz{\mathbf{z}}
		\global\long\def\mbZ{\mathbf{Z}}
		\global\long\def\barmbZ{\bar{\mathbf{Z}}}
		
		\global\long\def\hatmba{\widehat{\mathbf{a}}}
		\global\long\def\hatmbA{\widehat{\mathbf{A}}}
		\global\long\def\hatmbb{\widehat{\mathbf{b}}}
		\global\long\def\hatmbB{\widehat{\mathbf{B}}}
		\global\long\def\hatmbc{\widehat{\mathbf{c}}}
		\global\long\def\hatmbC{\widehat{\mathbf{C}}}
		\global\long\def\hatmbd{\widehat{\mathbf{d}}}
		\global\long\def\hatmbD{\widehat{\mathbf{D}}}
		\global\long\def\hatmbe{\widehat{\mathbf{e}}}
		\global\long\def\hatmbE{\widehat{\mathbf{E}}}
		\global\long\def\hatmbf{\widehat{\mathbf{f}}}
		\global\long\def\hatmbF{\widehat{\mathbf{F}}}
		\global\long\def\hatmbg{\widehat{\mathbf{g}}}
		\global\long\def\hatmbG{\widehat{\mathbf{G}}}
		\global\long\def\hatmbh{\widehat{\mathbf{h}}}
		\global\long\def\hatmbH{\widehat{\mathbf{H}}}
		\global\long\def\hatmbi{\widehat{\mathbf{i}}}
		\global\long\def\hatmbI{\widehat{\mathbf{I}}}
		\global\long\def\hatmbj{\widehat{\mathbf{j}}}
		\global\long\def\hatmbJ{\widehat{\mathbf{J}}}
		\global\long\def\hatmbk{\widehat{\mathbf{k}}}
		\global\long\def\hatmbK{\widehat{\mathbf{K}}}
		\global\long\def\hatmbl{\widehat{\mathbf{l}}}
		\global\long\def\hatmbL{\widehat{\mathbf{L}}}
		\global\long\def\hatmbm{\widehat{\mathbf{m}}}
		\global\long\def\hatmbM{\widehat{\mathbf{M}}}
		\global\long\def\hatmbn{\widehat{\mathbf{n}}}
		\global\long\def\hatmbN{\widehat{\mathbf{N}}}
		\global\long\def\hatmbo{\widehat{\mathbf{o}}}
		\global\long\def\hatmbO{\widehat{\mathbf{O}}}
		\global\long\def\hatmbp{\widehat{\mathbf{p}}}
		\global\long\def\hatmbP{\widehat{\mathbf{P}}}
		\global\long\def\hatmbq{\widehat{\mathbf{q}}}
		\global\long\def\hatmbQ{\widehat{\mathbf{Q}}}
		\global\long\def\hatmbr{\widehat{\mathbf{r}}}
		\global\long\def\hatmbR{\widehat{\mathbf{R}}}
		\global\long\def\hatmbs{\widehat{\mathbf{s}}}
		\global\long\def\hatmbS{\widehat{\mathbf{S}}}
		\global\long\def\hatmbt{\widehat{\mathbf{t}}}
		\global\long\def\hatmbT{\widehat{\mathbf{T}}}
		\global\long\def\hatmbu{\widehat{\mathbf{u}}}
		\global\long\def\hatmbU{\widehat{\mathbf{U}}}
		\global\long\def\hatmbv{\widehat{\mathbf{v}}}
		\global\long\def\hatmbV{\widehat{\mathbf{V}}}
		\global\long\def\hatmbw{\widehat{\mathbf{w}}}
		\global\long\def\hatmbW{\widehat{\mathbf{W}}}
		\global\long\def\hatmbx{\widehat{\mathbf{x}}}
		\global\long\def\hatmbX{\widehat{\mathbf{X}}}
		\global\long\def\hatmby{\widehat{\mathbf{y}}}
		\global\long\def\hatmbY{\widehat{\mathbf{Y}}}
		\global\long\def\hatmbz{\widehat{\mathbf{z}}}
		\global\long\def\hatmbZ{\widehat{\mathbf{Z}}}

		\global\long\def\tilmba{\widetilde{\mathbf{a}}}
		\global\long\def\tilmbA{\widetilde{\mathbf{A}}}
		\global\long\def\tilmbb{\widetilde{\mathbf{b}}}
		\global\long\def\tilmbB{\widetilde{\mathbf{B}}}
		\global\long\def\tilmbc{\widetilde{\mathbf{c}}}
		\global\long\def\tilmbC{\widetilde{\mathbf{C}}}
		\global\long\def\tilmbd{\widetilde{\mathbf{d}}}
		\global\long\def\tilmbD{\widetilde{\mathbf{D}}}
		\global\long\def\tilmbe{\widetilde{\mathbf{e}}}
		\global\long\def\tilmbE{\widetilde{\mathbf{E}}}
		\global\long\def\tilmbf{\widetilde{\mathbf{f}}}
		\global\long\def\tilmbF{\widetilde{\mathbf{F}}}
		\global\long\def\tilmbg{\widetilde{\mathbf{g}}}
		\global\long\def\tilmbG{\widetilde{\mathbf{G}}}
		\global\long\def\tilmbh{\widetilde{\mathbf{h}}}
		\global\long\def\tilmbH{\widetilde{\mathbf{H}}}
		\global\long\def\tilmbi{\widetilde{\mathbf{i}}}
		\global\long\def\tilmbI{\widetilde{\mathbf{I}}}
		\global\long\def\tilmbj{\widetilde{\mathbf{j}}}
		\global\long\def\tilmbJ{\widetilde{\mathbf{J}}}
		\global\long\def\tilmbk{\widetilde{\mathbf{k}}}
		\global\long\def\tilmbK{\widetilde{\mathbf{K}}}
		\global\long\def\tilmbl{\widetilde{\mathbf{l}}}
		\global\long\def\tilmbL{\widetilde{\mathbf{L}}}
		\global\long\def\tilmbm{\widetilde{\mathbf{m}}}
		\global\long\def\tilmbM{\widetilde{\mathbf{M}}}
		\global\long\def\tilmbn{\widetilde{\mathbf{n}}}
		\global\long\def\tilmbN{\widetilde{\mathbf{N}}}
		\global\long\def\tilmbo{\widetilde{\mathbf{o}}}
		\global\long\def\tilmbO{\widetilde{\mathbf{O}}}
		\global\long\def\tilmbp{\widetilde{\mathbf{p}}}
		\global\long\def\tilmbP{\widetilde{\mathbf{P}}}
		\global\long\def\tilmbq{\widetilde{\mathbf{q}}}
		\global\long\def\tilmbQ{\widetilde{\mathbf{Q}}}
		\global\long\def\tilmbr{\widetilde{\mathbf{r}}}
		\global\long\def\tilmbR{\widetilde{\mathbf{R}}}
		\global\long\def\tilmbs{\widetilde{\mathbf{s}}}
		\global\long\def\tilmbS{\widetilde{\mathbf{S}}}
		\global\long\def\tilmbt{\widetilde{\mathbf{t}}}
		\global\long\def\tilmbT{\widetilde{\mathbf{T}}}
		\global\long\def\tilmbu{\widetilde{\mathbf{u}}}
		\global\long\def\tilmbU{\widetilde{\mathbf{U}}}
		\global\long\def\tilmbv{\widetilde{\mathbf{v}}}
		\global\long\def\tilmbV{\widetilde{\mathbf{V}}}
		\global\long\def\tilmbw{\widetilde{\mathbf{w}}}
		\global\long\def\tilmbW{\widetilde{\mathbf{W}}}
		\global\long\def\tilmbx{\widetilde{\mathbf{x}}}
		\global\long\def\tilmbX{\widetilde{\mathbf{X}}}
		\global\long\def\tilmby{\widetilde{\mathbf{y}}}
		\global\long\def\tilmbY{\widetilde{\mathbf{Y}}}
		\global\long\def\tilmbz{\widetilde{\mathbf{z}}}
		\global\long\def\tilmbZ{\widetilde{\mathbf{Z}}}
		
		\renewcommand{\Vec}{\mathrm{Vec}}
		\newcommand{\bSigma}{{\bm \Sigma}}
		\newcommand{\bzero}{\mathbf{0}}
		
		\newcommand{\bX}{\mathbf{X}}
		\newcommand{\bY}{\mathbf{Y}}
		\newcommand{\bS}{\mathbf{S}}
		\newcommand{\bx}{\mathbf{x}}
		\newcommand{\bV}{\mathbf{V}}
		\newcommand{\bv}{\mathbf{v}}
		\newcommand{\bG}{\mathbf{G}}
		\newcommand{\bbeta}{{\bm\beta}}
		\newcommand{\bgamma}{{\bm\gamma}}
		\newcommand{\bGamma}{{\bm\Gamma}}
		\newcommand{\bLambda}{{\bm\Lambda}}
		
		\newcommand{\bA}{\mathbf{A}}
		\newcommand{\bB}{\mathbf{B}}
		\newcommand{\bC}{\mathbf{C}}
		\newcommand{\bD}{\mathbf{D}}
		\newcommand{\bI}{\mathbf{I}}
		\newcommand{\bJ}{\mathbf{J}}
		\newcommand{\bR}{\mathbf{R}}
		\newcommand{\bu}{\mathbf{u}}
		\newcommand{\bz}{\mathbf{z}}
		\newcommand{\f}{\mathbf{f}}
		\newcommand{\bzeta}{{\bm \zeta}}
		\newcommand{\btheta}{{\bm \theta}}
		\newcommand{\bP}{\mathbf{P}}
		\newcommand{\bT}{\mathbf{T}}
		\newcommand{\bZ}{\mathbf{Z}}
		\newcommand{\bepsilon}{\bm \epsilon}
		\newcommand{\bmu}{\bm \mu}
		\newcommand{\bxi}{\bm \xi}
		\newcommand{\bW}{\mathbf{W}}
		\newcommand{\bOmega}{\bm \Omega}
		\newcommand{\bU}{\mathbf{U}}
		\newcommand{\bTheta}{\bm \Theta}
		\newcommand{\bpi}{\bm \pi}
		\newcommand{\balpha}{\bm \alpha}
		\newcommand{\bdelta}{\bm \delta}
		\newcommand{\bF}{\mathbf{F}}

		\global\long\def\hatf{\widehat{f}}

		\global\long\def\bolalpha{\boldsymbol{\alpha}}
		\global\long\def\bolbeta{\boldsymbol{\beta}}
		\global\long\def\bolgamma{\boldsymbol{\gamma}}
		\global\long\def\boldelta{\boldsymbol{\delta}}
		\global\long\def\bolepsilon{\boldsymbol{\epsilon}}
		\global\long\def\bolzeta{\boldsymbol{\zeta}}
		\global\long\def\boleta{\boldsymbol{\eta}}
		\global\long\def\boltheta{\boldsymbol{\theta}}
		\global\long\def\bolkappa{\boldsymbol{\kappa}}
		\global\long\def\bollambda{\boldsymbol{\lambda}}
		\global\long\def\bolmu{\boldsymbol{\mu}}
		\global\long\def\bolnu{\boldsymbol{\nu}}
		\global\long\def\bolxi{\boldsymbol{\xi}}
		\global\long\def\bolpi{\boldsymbol{\pi}}
		\global\long\def\bolrho{\boldsymbol{\rho}}
		\global\long\def\bolsigma{\boldsymbol{\sigma}}
		\global\long\def\boltau{\boldsymbol{\tau}}
		\global\long\def\bolphi{\boldsymbol{\phi}}
		\global\long\def\bolchi{\boldsymbol{\chi}}
		\global\long\def\bolpsi{\boldsymbol{\psi}}
		\global\long\def\bolomega{\boldsymbol{\omega}}
		\global\long\def\bolGamma{\boldsymbol{\Gamma}}
		\global\long\def\bolDelta{\boldsymbol{\Delta}}
		\global\long\def\bolTheta{\boldsymbol{\Theta}}
		\global\long\def\bolLambda{\boldsymbol{\Lambda}}
		\global\long\def\bolPi{\boldsymbol{\Pi}}
		\global\long\def\bolSigma{\boldsymbol{\Sigma}}
		\global\long\def\bolPhi{\boldsymbol{\Phi}}
		\global\long\def\bolPsi{\boldsymbol{\Psi}}
		\global\long\def\bolOmega{\boldsymbol{\Omega}}

		\global\long\def\hatbolalpha{\widehat{\boldsymbol{\alpha}}}
		\global\long\def\hatbolbeta{\widehat{\boldsymbol{\beta}}}
		\global\long\def\hatbolgamma{\widehat{\boldsymbol{\gamma}}}
		\global\long\def\hatboldelta{\widehat{\boldsymbol{\delta}}}
		\global\long\def\hatbolepsilon{\widehat{\boldsymbol{\epsilon}}}
		\global\long\def\hatbolzeta{\widehat{\boldsymbol{\zeta}}}
		\global\long\def\hatboleta{\widehat{\boldsymbol{\eta}}}
		\global\long\def\hatboltheta{\widehat{\boldsymbol{\theta}}}
		\global\long\def\hatbolkappa{\widehat{\boldsymbol{\kappa}}}
		\global\long\def\hatbollambda{\widehat{\boldsymbol{\lambda}}}
		\global\long\def\hatbolmu{\widehat{\boldsymbol{\mu}}}
		\global\long\def\hatbolnu{\widehat{\boldsymbol{\nu}}}
		\global\long\def\hatbolxi{\widehat{\boldsymbol{\xi}}}
		\global\long\def\hatbolpi{\widehat{\boldsymbol{\pi}}}
		\global\long\def\hatbolrho{\widehat{\boldsymbol{\rho}}}
		\global\long\def\hatbolsigma{\widehat{\boldsymbol{\sigma}}}
		\global\long\def\hatboltau{\widehat{\boldsymbol{\tau}}}
		\global\long\def\hatbolphi{\widehat{\boldsymbol{\phi}}}
		\global\long\def\hatbolchi{\widehat{\boldsymbol{\chi}}}
		\global\long\def\hatbolpsi{\widehat{\boldsymbol{\psi}}}
		\global\long\def\hatbolomega{\widehat{\boldsymbol{\omega}}}
		\global\long\def\hatbolGamma{\widehat{\boldsymbol{\Gamma}}}
		\global\long\def\hatbolDelta{\widehat{\boldsymbol{\Delta}}}
		\global\long\def\hatbolTheta{\widehat{\boldsymbol{\Theta}}}
		\global\long\def\hatbolLambda{\widehat{\boldsymbol{\Lambda}}}
		\global\long\def\hatbolPi{\widehat{\boldsymbol{\Pi}}}
		\global\long\def\hatbolSigma{\widehat{\boldsymbol{\Sigma}}}
		\global\long\def\hatbolPhi{\widehat{\boldsymbol{\Phi}}}
		\global\long\def\hatbolPsi{\widehat{\boldsymbol{\Psi}}}
		\global\long\def\hatbolOmega{\widehat{\boldsymbol{\Omega}}}

		\global\long\def\tilbolalpha{\widetilde{\boldsymbol{\alpha}}}
		\global\long\def\tilbolbeta{\widetilde{\boldsymbol{\beta}}}
		\global\long\def\tilbolgamma{\widetilde{\boldsymbol{\gamma}}}
		\global\long\def\tilboldelta{\widetilde{\boldsymbol{\delta}}}
		\global\long\def\tilbolepsilon{\widetilde{\boldsymbol{\epsilon}}}
		\global\long\def\tilbolzeta{\widetilde{\boldsymbol{\zeta}}}
		\global\long\def\tilboleta{\widetilde{\boldsymbol{\eta}}}
		\global\long\def\tilboltheta{\widetilde{\boldsymbol{\theta}}}
		\global\long\def\tilbolkappa{\widetilde{\boldsymbol{\kappa}}}
		\global\long\def\tilbollambda{\widetilde{\boldsymbol{\lambda}}}
		\global\long\def\tilbolmu{\widetilde{\boldsymbol{\mu}}}
		\global\long\def\tilbolnu{\widetilde{\boldsymbol{\nu}}}
		\global\long\def\tilbolxi{\widetilde{\boldsymbol{\xi}}}
		\global\long\def\tilbolpi{\widetilde{\boldsymbol{\pi}}}
		\global\long\def\tilbolrho{\widetilde{\boldsymbol{\rho}}}
		\global\long\def\tilbolsigma{\widetilde{\boldsymbol{\sigma}}}
		\global\long\def\tilboltau{\widetilde{\boldsymbol{\tau}}}
		\global\long\def\tilbolphi{\widetilde{\boldsymbol{\phi}}}
		\global\long\def\tilbolchi{\widetilde{\boldsymbol{\chi}}}
		\global\long\def\tilbolpsi{\widetilde{\boldsymbol{\psi}}}
		\global\long\def\tilbolomega{\widetilde{\boldsymbol{\omega}}}
		\global\long\def\tilbolGamma{\widetilde{\boldsymbol{\Gamma}}}
		\global\long\def\tilbolDelta{\widetilde{\boldsymbol{\Delta}}}
		\global\long\def\tilbolTheta{\widetilde{\boldsymbol{\Theta}}}
		\global\long\def\tilbolLambda{\widetilde{\boldsymbol{\Lambda}}}
		\global\long\def\tilbolPi{\widetilde{\boldsymbol{\Pi}}}
		\global\long\def\tilbolSigma{\widetilde{\boldsymbol{\Sigma}}}
		\global\long\def\tilbolPhi{\widetilde{\boldsymbol{\Phi}}}
		\global\long\def\tilbolPsi{\widetilde{\boldsymbol{\Psi}}}
		\global\long\def\tilbolOmega{\widetilde{\boldsymbol{\Omega}}}

		\global\long\def\barbolmu{\overline{\bolmu}}
		\global\long\def\barmbX{\overline{\mbX}}

		\global\long\def\mbbR{\mathbb{R}}
		\global\long\def\mbbS{\mathbb{S}}
		\global\long\def\mbbX{\mathbb{X}}
		\global\long\def\mbbY{\mathbb{Y}}
		\global\long\def\mbbZ{\mathbb{Z}}
		\global\long\def\mbbU{\mathbb{U}}
		
		\global\long\def\calA{\mathcal{A}}
		\global\long\def\calB{\mathcal{B}}
		\global\long\def\calC{\mathcal{C}}
		\global\long\def\calD{\mathcal{D}}
		\global\long\def\calE{\mathcal{E}}
		\global\long\def\calF{\mathcal{F}}
		\global\long\def\calG{\mathcal{G}}
		\global\long\def\calH{\mathcal{H}}
		\global\long\def\calI{\mathcal{I}}
		\global\long\def\calJ{\mathcal{J}}
		\global\long\def\calK{\mathcal{K}}
		\global\long\def\calL{\mathcal{L}}
		\global\long\def\calM{\mathcal{M}}
		\global\long\def\calN{\mathcal{N}}
		\global\long\def\calO{\mathcal{O}}
		\global\long\def\calP{\mathcal{P}}
		\global\long\def\calQ{\mathcal{Q}}
		\global\long\def\calR{\mathcal{R}}
		\global\long\def\calS{\mathcal{S}}
		\global\long\def\calT{\mathcal{T}}
		\global\long\def\calU{\mathcal{U}}
		\global\long\def\calV{\mathcal{V}}
		\global\long\def\calW{\mathcal{W}}

		\global\long\def\mbell{\boldsymbol{\ell}}
		\global\long\def\bolell{\boldsymbol{\ell}}
		\global\long\def\mbzero{\mathbf{0}}

		\global\long\def\bolPhio{\boldsymbol{\Phi}_{0}}
		\global\long\def\bolOmegao{\boldsymbol{\Omega}_{0}}

		\global\long\def\bolSigmaX{\bolSigma_{\mbX}}
		\global\long\def\bolSigmaY{\bolSigma_{\mbY}}
		\global\long\def\bolSigmaXY{\boldsymbol{\Sigma}_{\mbX\mbY}}
		\global\long\def\mbSX{\mathbf{S}_{\mbX}}
		\global\long\def\mbSY{\mathbf{S}_{\mbY}}
		\global\long\def\mbSXY{\mathbf{S}_{\mbX\mbY}}
		\global\long\def\mbSYX{\mathbf{S}_{\mbY\mbX}}
		\global\long\def\mbRYX{\mathbf{S}_{\mbY|\mbX}}
		\global\long\def\mbRXY{\mathbf{S}_{\mbX|\mbY}}
		\global\long\def\mbSc{\mbS_{\mbC}}
		\global\long\def\mbSd{\mbS_{\mbD}}

		\global\long\def\sumn{\sum_{i=1}^{n}}

		\global\long\def\E{\mathrm{E}}
		\global\long\def\F{\mathrm{F}}
		\global\long\def\J{\mathrm{J}}
		\global\long\def\H{\mathrm{H}}
		\global\long\def\G{\mathrm{G}}
		\global\long\def\Cov{\mathrm{cov}}
		\global\long\def\Corr{\mathrm{corr}}
		\global\long\def\Var{\mathrm{var}}
		\global\long\def\dimension{\mathrm{dim}}
		\global\long\def\spn{\mathrm{span}}
		\global\long\def\vech{\mathrm{vech}}
		\global\long\def\vecc{\mathrm{vec}}
		\global\long\def\Prob{\mathrm{Pr}}
		\global\long\def\Env{\mathrm{env}}
		\global\long\def\tr{\mathrm{tr}}
		\global\long\def\dg{\mathrm{diag}}
		\global\long\def\asyVar{\mathrm{avar}}
		\global\long\def\MSE{\mathrm{MSE}}
		\global\long\def\OLS{\mathrm{OLS}}

		\global\long\def\sigerr{\sigma_{e}^{2}}
		\global\long\def\hatsigerr{\widehat{\sigma}_{e}^{2}}
		\global\long\def\bolSigmaf{\bolSigma_{\mbf}}
		\global\long\def\tilssigma{\widetilde{\sigma}^{2}}
		\global\long\def\hatssigma{\widehat{\sigma}^{2}}
		\global\long\def\ssigma{\sigma^{2}}
		\global\long\def\PLS{\mathrm{PLS}}
		\global\long\def\hatlambda{\widehat{\lambda}}
		\global\long\def\hatpi{\widehat{\pi}}
		
		\global\long\def\CRE{\mathcal{R}_{\bolSigma_{Y}}(\calB)}
		
		\global\long\def\CS{\calS_{Y\mid\mbX}}
		\global\long\def\hatsigma{\widehat{\sigma}}
		\global\long\def\hatdelta{\widehat{\delta}}
		\global\long\def\hatb{\widehat{b}}
		\global\long\def\tilb{\widetilde{b}}
		\global\long\def\hatgamma{\widehat{\gamma}}
		\global\long\def\hatpi{\widehat{\pi}}

		\newtheorem{lemma}{Lemma}
		\newtheorem{proposition}{Proposition}
		\newtheorem{theorem}{Theorem}
		\newtheorem{definition}{Definition}
		\newtheorem{example}{Example}
		
		\newcommand{\beqn}{\begin{equation*}}
		\newcommand{\eeqn}{\end{equation*}}
		
		\newcommand{\bea}{\begin{eqnarray}}
		\newcommand{\eea}{\end{eqnarray}}
		
		\newcommand{\bean}{\begin{eqnarray*}}
			\newcommand{\eean}{\end{eqnarray*}}
		
		\newcommand{\beq}{\begin{equation}}
		\newcommand{\eeq}{\end{equation}}
		
		\newcommand{\texthl}[1]{{ {\color{blue} #1}}}
				\global\long\def\supB{\sup_{\boltheta\in{\cal B}_{con}}}

\title{A Doubly-Enhanced EM Algorithm for\\ Model-Based Tensor Clustering\thanks{Corresponding author: Xin Zhang (xzhang8@fsu.edu). The authors would like to thank the Co-Editors, Associate Editor and reviewers for helpful comments. Research for this paper was supported in part by grants CCF-1617691 and CCF-1908969 from the National Science Foundation.  } }
\author{
\bigskip
Qing Mai, Xin Zhang, Yuqing Pan and Kai Deng\\
\textit{Florida State University}
}
\date{}
\maketitle

\begin{abstract}
Modern scientific studies often collect data sets in the form of tensors. These datasets call for innovative statistical analysis methods. 
In particular, there is a pressing need for tensor clustering methods to understand the heterogeneity in the data. 
We propose a tensor normal mixture model approach to enable probabilistic interpretation and computational tractability.
Our statistical model leverages the tensor covariance structure to reduce the number of parameters for parsimonious modeling, and at the same time explicitly exploits the correlations for better variable selection and clustering. 
We propose a doubly-enhanced expectation-maximization (DEEM) algorithm to perform clustering under this model. 
Both the Expectation-step and the Maximization-step are carefully tailored for tensor data in order to maximize statistical accuracy and minimize computational costs in high dimensions. 
Theoretical studies confirm that DEEM achieves consistent clustering even when the dimension of each mode of the tensors grows at an exponential rate of the sample size. Numerical studies demonstrate favorable performance of DEEM in comparison to existing methods.\\

{\centering
\textbf{Keywords:} Clustering; the EM Algorithm; Gaussian Mixture Models; Kronecker Product Covariance; Minimax; Tensor.
}

\end{abstract}

\newpage

\section{Introduction}\label{Sec: Intro}

Tensor data are increasingly popular in modern scientific studies. Research in brain image analysis, personalized recommendation and multi-tissue multi-omics studies often collect data in the form of matrices (i.e, 2-way tensors) or higher-order tensors for each observation. The tensor structure brings challenges to the statistical analysis. On one hand, tensor data are often naturally high-dimensional. This leads to an excessive number of parameters in statistical modeling. On the other hand, the tensor structure contains information that cannot be easily exploited by classical multivariate, i.e.~vector-based, methods. Motivated by the prevalence of tensor data and the challenges to statistical analysis, a large number of novel tensor-based methods have been developed in recent years. There is a rapidly growing literature on the analysis of tensor data, for example, on tensor decomposition \citep{ chi2012tensors,sun2016provable, zhang2017guaranteed},  regression \citep{zhou2013tensor, hoff2015multilinear, raskutti2015convex, wang2016generalized, LiZhang2015, zhang2016tensor, lock2017tensor} and classification \citep{tensorDWD,CATCH}. These methods, among many others, take advantage of the tensor structure to drastically reduce the number of parameters, and use tensor algebra to streamline estimation and advance theory. 

We study the problem of model-based tensor clustering. When datasets are heterogeneous, cluster analysis sheds light on the heterogeneity by grouping observations into clusters such that observations within each cluster are similar to each other, but there is noticeable difference among clusters. For more background, see \citet{review1} and \citet{review3} for  overviews of model-based clustering. 
Various approaches have been proposed in recent years for clustering on high-dimensional vector data \citep{Ng2001,Law2004,Arthur2007, Pan2007,Wang2008,Guo2010, Witten2010SKM,  Cai2017CHIME,verzelen2017,Hao2018}. Although many of these vector methods could be applied to tensor data by vectorizing the tensors first, this brute-force approach is generally not recommended, because the vectorization completely ignores the tensor structure. As a result, vectorization could often lead to loss of information, and thus efficiency and accuracy. It is much more desirable to have clustering methods specially designed for tensor data.

Model-based clustering often assumes a finite mixture of distributions for the data. In particular, the Gaussian mixture model  (GMM) plays an important role in high-dimensional statistics due to its flexibility, interpretability and computational convenience. 
Motivated by GMM, we consider  a tensor normal mixture model (TNMM). In comparison to the existing GMM methods for vector data,  TNMM exploits the tensor covariance structure to drastically reduce the total number of parameters in covariance modeling. 
Thanks to the simplicity of matrix/tensor normal distributions, clustering and parameter estimation is straightforward based on the expectation-maximization (EM) algorithm \citep{Dempster1977} . Among others, 
\citet{viroli2011finite,anderlucci2015covariance,Gao2018,gallaugher2018finite} are all extensions of GMM from vector to matrix, but are not directly applicable to higher-order tensors. Moreover, the focus of these works is computation and applications in the presence of additional information, such as covariates, longitudinal correlation, heavy tails and skewness in the data, but no theoretical results are provided for high dimensional data analysis. 
The GMMs can be straightforwardly extended to higher-order tensors adopting the standard EM algorithm. 
However, as we demonstrate in numerical studies, the standard EM can be dramatically improved by our Doubly-Enhanced Expectation-Maximization (DEEM) algorithm.
 
The DEEM algorithm is developed under TNMM to efficiently incorporate tensor correlation structure and variable selection for clustering and parameter estimation. Similar to classical EM algorithms, DEEM iteratively carries out an enhanced E-step and an enhanced M-step. In the enhanced E-step, we impose sparsity directly on the optimal clustering rule as a flexible alternative to popular low-rank assumptions on tensor coefficients. The variable selection empowers DEEM to high-dimensional tensor data analysis. In the enhanced M-step, we employ a new estimator for the tensor correlation structure, which facilitates both the computation and the theoretical studies. These modifications to the standard EM algorithm are very intuitive and practically motivated. More importantly, we show that the clustering error of DEEM converges to the optimal clustering error at the minimax optimal rate. DEEM is also highly competitive in empirical studies. 

To achieve variable selection and clustering simultaneously, we impose the sparsity assumption on our model and then incorporate a penalized estimator in DEEM. Although penalized estimation is a common strategy in high-dimensional clustering, there are many different approaches. For example , \citet{Wang2008} penalize cluster means;  \citet{Guo2010, verzelen2017} penalize cluster mean differences; \citet{Pan2007, Witten2010SKM, Law2004} achieve variable selection by assuming independence among variables; \citet{Hao2018} impose sparsity on both cluster means and precision matrices; \citet{Cai2017CHIME} impose sparsity on the discriminant vector. Our approach is similar to \citet{Cai2017CHIME} in that our sparsity assumption is directly imposed on the discriminant tensor coefficients -- essentially a re-parameterization of the means and covariance matrices to form sufficient statistics in clustering. As a result of this parameterization, the correlations among variables are utilized in variable selection, while the parameter of interest has the same dimensionality as the cluster mean difference. 

Due to the non-convex nature of cluster analysis, conditions on the initial value are commonly imposed in theoretical studies. Finding theoretically guaranteed initial values for cluster analysis is an important research area on its own, with many interesting works under GMM \citep{kalai2010efficiently,moitra2010settling,hsu2013learning,hardt2015tight}. 
To provide a firmer theoretical ground for the consistency of DEEM, we further develop an initialization algorithm for TNMM in general, which may be of independent interest. A brief discussion on the initialization is provided in Section~\ref{Sec: C1}. The detailed algorithm (Algorithm~S.4) and related theoretical studies are provided in Section~G of Supplementary Materials.

Two related but considerably different problems are worth-mentioning, but beyond the scope of this article. The first is the low-rank approximation in K-means clustering \citep{macqueen1967, cohen2015dimensionality}. For example, \citet{Sun2018} use tensor decomposition in the minimization of the total squared Euclidean distance of each observation to its cluster centroid. 
While the low-rank approximation is widely adopted in tensor data analysis, our method is more directly targeted at the optimal rule of clustering under the TNMM, and does not require low-rank structure of the tensor coefficients.
The second is the clustering of features (variables) instead of, or, along with observations.
Clustering variables into similar groups has applications in a wide range of areas such as genetics, text mining and imaging analysis, and also has attracted substantial interest in theoretical studies. For example,
\citet{bing2019adaptive,bunea2020model} studied feature clustering in high dimensions; \citet{lee2010biclustering,tan2014sparse,chi2017convex} developed bi-clustering methods that simultaneously group features and observations into clusters. 
Extensions of the feature-sample bi-clustering for vector observations are known as the \emph{co-clustering} or \emph{multiway clustering} problems \citep{kolda2008scalable,jegelka2009approximation,chi2018provable,wang2019multiway}, where each mode of the tensor is clustered into groups, resulting in a checkerbox structure. Our problem is different from these works in that our sole goal is to cluster the observations.  

The rest of the paper is organized as follows. In Section~\ref{sec: model}, we formally introduce the model and discuss the importance of modeling the correlation structure. In Section~\ref{Sec:DEEM}, we propose the DEEM algorithm. Theoretical results are presented in Section~\ref{Sec: theory}. Section~\ref{Sec: num} contains numerical studies on simulated and real data. Additional numerical studies, proofs and other technical details are relegated to Supplementary Materials.

\section{The Model} \label{sec: model}

\subsection{Notation and Preliminaries}\label{Sec: notation}

A multi-dimensional array $\mbA\in\mbbR^{p_1\times\cdots\times p_M}$ is called an $M$-way tensor. We denote ${\cal J}=(j_1,\ldots,j_M)$ as the index of one element in the tensor. 
The vectorization of $\mbA$ is a vector, $\vecc(\mbA)$, of length $(\prod_{m=1}^Mp_m)$. The mode-$k$ matricization of a tensor is a matrix of dimension $(p_k\times\prod_{m\neq k}p_m)$, denoted by $\mbA_{(k)}$, where the $(j_1,\ldots,j_M)$-th element of $\mbA$ is the $(j_k,l)$-th element of $\mbA_{(k)}$ with $l=1+\sum_{m=1,m\ne k}^M\{(j_m-1)\prod_{t=1,t\ne k}^{m-1}p_t\}$. A tensor $\mbC\in\mbbR^{d_1\times\cdots\times d_M}$ can be multiplied with a $d_m\times p_m$ matrix $\mbG_m$ on the $m$-th mode, denoted as $\mbC\times_m\mbG_m\in\mathbb{R}^{d_1\times\cdots\times d_{m-1}\times p_m\times d_{m+1}\times\cdots\times d_M}$. If $\mbA=\mbC\times_1\mbG_1\times\cdots\times_M \mbG_M$, we equivalently write the Tucker decomposition of $\mbA$ as $\mbA=\llbracket \mbC;\mbG_1,\ldots,\mbG_M\rrbracket$. A useful fact is that $\vecc(\llbracket \mbC;\mbG_1,\ldots,\mbG_M\rrbracket)=(\mbG_M\otimes\cdots\otimes \mbG_1)\vecc(\mbC)\equiv (\bigotimes_{m=M}^{m=1}\mbG_m)\vecc(\mbC)$, where $\otimes$ represents the Kronecker product. The inner product of two tensors $\mbA,\mbB$ of matching dimensions is defined as $\langle \mbA,\mbB\rangle=\sum_{{\cal J}} a_{\cal J}b_{\cal J}$. For more background on tensor algebra, see \citet{KoldaBader09Tensor}.

The tensor normal distribution is an extension of matrix multivariate normal distribution \citep{gupta1999matrix, hoff2011separable}. For a random tensor $\mbX\in\mathbb{R}^{p_1\times\cdots\times p_M}$, if $\mbX=\bolmu+\llbracket \mbZ;\bolSigma_1^{1/2},\ldots,\bolSigma_M^{1/2}\rrbracket$ for $\bolmu\in\mathbb{R}^{p_1\times\cdots \times p_M},\bolSigma_m\in\mathbb{R}^{p_m\times p_m},$ and $Z_{\cal J}{\sim}N(0,1)$ independently, we say that $\mbX$ follows the tensor normal distribution. We often use the shorthand notation $\mbX\sim TN(\bolmu;\bolSigma_1,\ldots,\bolSigma_M)$. Because $\vecc(\mbX)=\vecc(\llbracket \mbZ;\bolSigma_1,\ldots,\bolSigma_M)\rrbracket=(\bigotimes_{m=M}^{m=1}\bolSigma_m)\vecc(\mbZ)$, we have that $\mbX\sim TN(\bolmu;\bolSigma_1,\ldots,\bolSigma_M)$ if $\vecc(\mbX)\sim N(\vecc(\bolmu),\bigotimes_{m=M}^{m=1}\bolSigma_m)$.
The parameters $\bolSigma_1,\ldots,\bolSigma_M$ are only identifiable up to $M$ rescaling constants. For example, for any set of positive constants $g_1,\ldots, g_M$ such that $\prod_{m=1}^{M}g_m=1$, we have $\bigotimes_{m=M}^{m=1} (g_m\bolSigma_m)=\bigotimes_{m=M}^{m=1}\bolSigma_m$. It is then easy to verify that $TN(\bolmu;g_1\bolSigma_1,\ldots,g_M\bolSigma_M)$ is the same distribution as $TN(\bolmu;\bolSigma_1,\ldots,\bolSigma_M)$.

We next briefly review the Gaussian mixture model \citep[GMM,][]{banfield1993model}. The GMM with shared covariance assumes that 
observations $\mbU_i\in\mathbb{R}^p$, $i=1,\ldots, n$, are independent and identically distributed (i.i.d.) with the mixture normal distribution
$\sum_{k=1}^K \pi_k^*N(\bolphi_k^*,\bolPsi^*)$,
where $K$ is a positive integer, $\pi_k^*\in (0,1)$ is the prior probability for the $k$-th cluster, $\bolphi_k^*\in\mathbb{R}^p$ is the cluster mean within the $k$-th cluster, and the symmetric positive definite matrix $\bolPsi^*\in\mathbb{R}^{p\times p}$ is the within-cluster covariance.  We note that the within-cluster covariance could be different across clusters. But we choose to present GMM with constant within-cluster covariance, because it is more closely related to our study.
The latent cluster representation of the GMM is often used to connect it with discriminant analysis, optimal clustering rules, and the EM algorithm. 
Specifically, the GMM can be written equivalently as
\beq\label{eq.GMM}
\Pr(Y_i=k)=\pi_k^*,\quad \mbU_i\mid (Y_i=k) \sim N(\bolphi_k^*,\bolPsi^*),
\eeq
where the latent variables $Y_i\in\{1,\ldots, K\}$. We use the superscript $^*$ to denote the true value of a parameter in population.

\subsection{The Tensor Normal Mixture Model} \label{sec: clustermodel}

Consider independent tensor-variate observations $\mbX_i\in\mathbb{R}^{p_1\times\cdots \times p_M}$, $i=1,\ldots,n$. The observations are heterogeneous in that they are drawn from $K$ clusters, but the cluster labels are unavailable to us. To recover these labels, we assume that $\mbX_i$ follows a mixture of tensor normal (TN) distributions (cf. Section~\ref{Sec: notation}) such that,
\begin{equation}\label{margdis}
\mbX_i\sim\sum_{k=1}^K\pi_k^*TN(\bolmu_k^*;\bolSigma_1^*,\ldots,\bolSigma_M^*),\quad i=1,\dots,n,
\end{equation}
where $\bolmu_k^*\in\mathbb{R}^{p_1\times \cdots\times p_M}$ is the mean of the $k$-th cluster, $\bolSigma_m^*\in\mathbb{R}^{p_m\times p_m}$ is the common within-class covariance along mode $m$, and $0<\pi_k^*<1$ is the prior probability for $\mbX_i$ to be in the $k$-th cluster such that $\sum_{k=1}^K \pi_k^*=1$. 
Throughout the rest of this paper, we use $\sigma^*_{m,ij}$ to denote the $(i,j)$-th entry in $\bolSigma^*_m$. To ensure the identifiability of the covariance matrices, we assume that $\sigma^*_{m,11}=1$ for $m>1$, and $\sigma^*_{1,11}$ is the variance of $X_{i,1\cdots 1}$ within clusters, {where $X_{i,1\cdots 1}$ is the $(1,\cdots,1)$-th element in the tensor $\mbX_{i}$}. We will explicitly specify the scale of $\bolSigma^*_1$ shortly. We refer to the model \eqref{margdis} as the tensor normal mixture model (TNMM). 

Parallel to the latent variable representation in GMM, \eqref{eq.GMM}, we introduce the latent cluster membership $Y_i\in\{1,\dots,K\}$ and re-write \eqref{margdis} as
\begin{equation}\label{modellatent}
\text{Pr}(Y_i=k)=\pi_k^*,\quad \mbX_i\mid (Y_i=k)\sim TN(\bolmu_k^*;\bolSigma^*_1,\ldots,\bolSigma^*_M).
\end{equation}
Intuitively, the TNMM assumes that $\mbX_i$ follows a tensor normal distribution with mean $\bolmu_k^*$ within the $k$-th cluster.  The parameter $\bolmu_k^*$ represents the centroid of the $k$-th cluster, while the covariance matrices $\bolSigma_1^*,\ldots,\bolSigma_M^*$ determine the dependence structure among the features. Also, with the latent variable $Y_i$, it is easy to specify the scale of $\bolSigma_1^*$. Since $\sigma^*_{m,11}=1$ for all $m>1$, we must have that $\sigma^*_{1,11}=\Var(X_{i,1\cdots 1}\mid Y_i=k)$ for all $i,k$.

To better understand the TNMM, we consider its implication on $\vecc(\mbX_i)$. By vectorizing the data, the model is equivalent to
\beq\label{eq.TNMM.vec}
\Pr(Y_i=k)=\pi_k^*,\quad\vecc(\mbX_i)\mid (Y_i=k)\sim N(\vecc(\bolmu^*_k),\bigotimes_{m=M}^1\bolSigma^*_m),
\eeq
which resembles the GMM in \eqref{eq.GMM}. A major distinction arises from our parsimonious parametrization of the covariance. It is easy to see that, if we ignore the tensor structure and impose GMM on $\vecc(\mbX_i)$, the covariance has $O(\prod_{i=1}^M p_m^2)$ parameters. However, the covariance in \eqref{eq.TNMM.vec} is determined by $O(\sum_{m=1}^Mp_m^2)$ parameters, because of the separable Kronecker product structure in the tensor covariance. The reduction in the number of parameters is drastic even for moderately high dimensions, and improves estimation efficiency, especially when the sample size is small. 

Note that the vectorization is only for demonstration purpose. In our estimation algorithm to be introduced, we never vectorize the observations; instead, we preserve the tensor form and use tensor operators for efficient implementation. When it comes to methodology developments and computation, the tensor form also greatly reduces the storage and computation costs in the DEEM algorithm. See Section~\ref{Sec: benefit.tensor} for details.


Many existing methods for tensor data analysis employ the tensor normal assumption in seek of parsimony and simplicity in likelihood-based procedure \citep{hoff2011separable,fosdick2014separable,LiZhang2015,CATCH}. Such an assumption has demonstrated success in regression and classification problems, which motivates the application of TNMM to unsupervised tensor learning. On the other hand, although it is known in low dimensions that modeling the dependence benefits clustering, many high-dimensional clustering methods ignore the correlation structure in data when performing variable selection and dimension reduction. 
It makes intuitive sense that modeling the correlation continues to improve clustering accuracy in high dimensions, but careful investigation further reveals that correlations heavily impact the variable selection as well. We explain this point in the next section. 



\subsection{Optimal clustering rule and variable selection}\label{sec: cor}

Many methods in the literature ignore the correlations among features when performing clustering in high dimensions. Some of them are developed based on the K-means clustering, and hence make no attempt to model the correlations \citep[e.g]{Witten2010SKM,Sun2018, Cao2013}; others  assume that the features are independent within each cluster and thus eliminate the need to model the correlations \citep[e.g]{Pan2007,Guo2010}. For simplicity, we refer to methods that ignore the correlation structure among features as independence methods. We demonstrate the impacts of correlations on variable selection by comparing the target clustering rule of independence methods to the optimal clustering rule under the TNMM \eqref{margdis}.

Consider $\mbX$ with conditional probability density function $f_k$ within the $k$-th cluster. The optimal classification rule defined on the population level is
\beq\label{eq.opt.generic}
\widehat{Y}^{opt}= \phi^{Bayes}(\mbX)= \arg\max_k\pi_k^*f_k(\mbX),
\eeq
where $\pi_k^*$ is the marginal probability for $\mbX$ to belong to the $k$-th cluster.
Although the above rule in \eqref{eq.opt.generic} is commonly known as the Bayes rule for classification, it continues to be optimal for clustering. 

First of all, we define the clustering error of a population (non-stochastic) classifier $\phi$ as $\min_{\Pi}\Pr(\phi(\mbX)\neq \Pi(Y) )$, where we optimize over all possible permutations of the $K$ labels $\Pi: \{1,\dots,K\}\mapsto \{1,\dots,K\}$. A major difference between classification and clustering problems is that the $K$ labels are well-defined in classification but are artificially created in clustering. As a result of this completely latent and non-identifiable cluster labels, any clustering rule $\phi(\mbX): \mbbR^{p_1\times\dots\times p_M}\mapsto \{1,\dots,K\}$ is equivalent to the permuted $\Pi\{\phi(\mbX)\}$. 

It is well-known that the Bayes rule $\phi^{Bayes}(\mbX)$ minimizes classification error. Because $\phi^{Bayes}(\mbX)$ produces a prediction that is solely based on $\mbX$ regardless of whether we observe $Y$ or not, it also minimizes the clustering error.
The rule defined in \eqref{eq.opt.generic} is thus optimal and is the target of our analysis. 
In estimation, the additional permutation operator needs to be carefully accounted for, making the clustering analysis much more challenging than classification. 


Recall that $X_{\cal J}$ is the ${\cal J}$-th element of $\mbX$, where ${\cal J}=(j_1,\ldots,j_M)$. 
For ease of presentation, we consider the special case of $K=2$ and $\diag(\bolSigma^*_m)=1$ for all $m$ throughout the rest of this section. 
Under the TNMM \eqref{margdis}, the optimal rule \eqref{eq.opt.generic} is equivalent to assigning $\mbX$ to Cluster 2 if and only if 
\beq\label{eq.opt}
\log{(\pi_2^*/\pi_1^*)}+\langle\mbX-\frac{\bolmu^*_1+\bolmu^*_2}{2},\mbB^*\rangle>0,
\eeq
where $\mbB^*=\llbracket\bolmu^*_2-\bolmu^*_1;(\bolSigma^*_1)^{-1},\ldots,(\bolSigma^*_M)^{-1}\rrbracket$. Consequently, an element $X_{\cal J}$ is not important for clustering if and only if $b^*_{{\cal J}}=0$. To achieve optimal clustering, we only need the variables in ${\cal D}=\{{\cal J}: b^*_{{\cal J}}\ne 0\}$. 

However, if we treat the variables as independent within each cluster, e.g.~as in many existing high-dimensional clustering methods, it is equivalent to assuming that $\bolSigma_m^*$ are all identity matrices under the TNMM. Then \eqref{eq.opt.generic} leads to the following ``independence rule'':
\beq
\widehat Y^{indep}=\arg\min_k\{-2\log\pi_k^*+ \sum_{{\cal J}}(X_{\cal J}-\mu^*_{k,{\cal J}})^2\}.
\eeq 
That is, $\widehat Y^{indep}=2$ if and only if $\log{(\pi_2^*/\pi_1^*)}+\langle\mbX-\frac{\bolmu^*_1+\bolmu^*_2}{2},\bolmu^*_2-\bolmu^*_1\rangle>0$.
Hence, the variable selection of the independence methods essentially targets at the set ${\cal A}=\{{\cal J}: \mu^*_{1,{\cal J}}\ne \mu^*_{2,{\cal J}}\}$.  

{It can be seen that the optimal rule in \eqref{eq.opt} is usually different from the independence rule, because $\mbB^*\ne \bolmu_2^*-\bolmu_1^*$ in general. Consequently, the independence methods can not achieve the optimal error rate when  the covariance matrices $\bolSigma_m^*$ are not diagonal.
Moreover, the difference between ${\cal A}$ and ${\cal D}$ implies that the correlation structure also impacts the variable selection results. Since $\mbB^*$ is a product between $\bolmu_2^*-\bolmu_1^*$ and $(\bolSigma_m^*)^{-1}, m=1,\ldots,M$, elements with constant means across clusters (i.e, elements in ${\cal A}^c$) could still improve clustering accuracy if they are correlated with $\mbX_{\cal A}$. In contrast, elements in ${\cal A}$ are not necessarily relevant for clustering, because their corresponding $b^*_{\cal J}$ could be zero. In Section~A of the Supplementary Materials, we construct examples to illustrate this phenomenon. {A similar discussion is available in \citet{DSDA} for discriminant analysis on vector data. But, to the best of our knowledge, we are the first to discuss this point for clustering on tensor data.}}

{Similar to the independence rule, K-means methods may also suffer from ignoring the correlations. Although (sparse) K-means can be viewed as model-free clustering methods, their target set for variable selection is similar to $\cal A$.  K-means clustering \citep[see Equation~(14.33) in][]{FHT01} searches for
$\arg\min_{\{Y_i\}_{i=1}^n,\{\mu_k\}_{k=1}^K}\sum_{k=1}^K n_k\sum_{Y_i=k}\sum_{\cal J}(X_{i,\cal J}-\mu_{k,{\cal J}})^2$, where $n_k$ is the size of the $k$-th cluster. Hence, if a feature has constant mean across clusters, it is not important in the final clustering. We only need the set of variables with different means, which resembles $\cal A$. }

\section{The doubly-enhanced EM algorithm}\label{Sec:DEEM}

We develop a general estimation procedure for TNMM \eqref{margdis} with $K\ge 2$, where we assume that $K$ is known. The clustering rule is directly obtained by plugging in the estimates of model parameters to the (population) optimal rule \eqref{eq.opt}. We first describe the standard EM algorithm in Section~\ref{Sec:EM.issues}. We further discuss the limitations of the standard EM that render it unsuitable for high-dimensional tensor clustering. 
Then we proceed to develop our DEEM algorithm and discuss its characteristics.

\subsection{The standard EM algorithm}\label{Sec:EM.issues}

The EM algorithm \citep{Dempster1977} is widely used in model-based clustering. Although we argue that the standard EM is not suitable for high-dimensional tensor clustering, it is nevertheless an inspiration of our DEEM algorithm and applicable in low-dimensional settings. We discuss the standard EM algorithm in what follows.

Define $\boltheta=\{\pi_k,\bolmu_k,k=1,\ldots,K; \bolSigma_m,m=1,\ldots, M\}$ as the model parameters in TNMM. Let $f(y,\mbx; \boltheta)$ denote the joint probability function of $Y$ and $\mbX$. If we could observe the latent variables $\{Y_i\}_{i=1}^n$, then the log-likelihood function for the complete data is  
\begin{equation}\label{loglikelihoodvec}
l_n(\boltheta)=\sum_{i=1}^n\log f(Y_i,\mbX_i; \boltheta)=\sum_{i=1}^n\{\log{\pi_{Y_i}}+\log f_{Y_i}(\mbX_i;\boltheta)\},
\end{equation}
where $f_{Y_i}(\mbX_i;\boltheta)$ is the conditional density function of $\mbX_i\mid Y_i$. From the tensor normal distribution, we have
\begin{equation}\label{classdensity}
 f_k(\mbX_i;\boltheta)=\frac{\exp(-\frac{1}{2}\langle \llbracket \mbX_i-\bolmu_k;\bolSigma_1^{-1},\ldots,\bolSigma_M^{-1}\rrbracket, \mbX_i-\bolmu_k\rangle)}{(2\pi)^{p/2}\lvert \bolSigma_1\rvert^{q_1/2}\cdots\lvert\bolSigma_M\rvert^{q_{M}/2}},
\end{equation} 
where $p=\prod_{m=1}^Mp_m$ and $q_m={p}/{p_m}$.

Clearly, the latent variables $\{Y_i\}_{i=1}^n$ are unobservable and thus we cannot directly maximize the log-likelihood function \eqref{loglikelihoodvec} to obtain the estimator of $\boltheta$. The EM algorithm tries to maximize $l_n(\boltheta)$ by iteratively performing the Expectation-step (E-step) and the Maximization-step (M-step).


Consider the $(t+1)$-th iteration with the current value $\widetilde{\boltheta}^{(t)}$. In the E-step, we evaluate
\beq\label{eq.Q.function}
Q_n(\boltheta\mid\widetilde{\boltheta}^{(t)})=\E\left[ l_n(\boltheta)\mid \{\mbX_i \}_{i=1}^{n},\widetilde{\boltheta}^{(t)}\right]=\sum_{i=1}^n\sum_{k=1}^K\widetilde{\xi}_{ik}^{(t)}\{\log{\pi_{k}}+\log f_{k}(\mbX_i;\boltheta)\},
\eeq
where 
\beq
\widetilde{\xi}_{ik}^{(t)}=\Pr(Y_i=k\mid\mbX_i,\widetilde{\boltheta}^{(t)})=\frac{\widetilde{\pi}^{(t)}_kf_k(\mbX_i;\widetilde{\boltheta}^{(t)})}{\sum_{j=1}^K \widetilde{\pi}^{(t)}_jf_j(\mbX_i;\widetilde{\boltheta}^{(t)})}.
\eeq
In the M-step, we maximize $Q_n(\boltheta\mid\widetilde{\boltheta}^{(t)})$ over $\boltheta$. The updates for $\pi_k$ and $\bolmu_k$ can be easily computed with an explicit form. However, the updates for $\bolSigma_1,\ldots,\bolSigma_M$ are much more difficult to obtain. With some calculation, we have the following lemma. 
 
\begin{lemma}\label{lem.standard.EM} 
 The maximizers of  \eqref{eq.Q.function} must satisfy
\begin{equation}
			\tilbolSigma_m^{(t+1)}=(nq_m)^{-1}{\sum_{i=1}^n\sum_{k=1}^K\widetilde{\xi}_{ik}^{(t+1)}\{\widetilde{\mbW}_{ik}^{(t+1)}\}\{\widetilde{\mbW}_{ik}^{(t+1)}\}\T},
			\end{equation}
			where $\widetilde{\mbW}_{ik}^{(t+1)}$ is the mode-$m$ matricization of the product
			\begin{equation}
			\llbracket \mbX_i-\widetilde\bolmu_k^{(t+1)}; \{\widetilde\bolSigma_1^{(t+1)}\}^{-\frac{1}{2}},\ldots,\{\widetilde\bolSigma_{m-1}^{(t+1)}\}^{-\frac{1}{2}},\mbI_{p_m},\{\widetilde\bolSigma_{m+1}^{(t+1)}\}^{-\frac{1}{2}},\ldots,\{\widetilde\bolSigma_M^{(t+1)}\}^{-\frac{1}{2}}\rrbracket.
			\end{equation}
\end{lemma}

Lemma~\ref{lem.standard.EM} implies that an iterative algorithm is needed to find $\widetilde{\bolSigma}_m^{(t+1)}$. Since $\widetilde{\bolSigma}^{(t+1)}_m$ depends on all the other covariance estimates $\widetilde{\bolSigma}_{m'}^{(t+1)},m'\ne m$, we need to update one covariance estimate while keeping all the others fixed until convergence to find $\widetilde{\bolSigma}_m^{(t+1)}$. By letting $K=1$, the results in Lemma~\ref{lem.standard.EM} also reproduce the maximum likelihood estimation in the tensor normal distribution \citep[e.g.][]{dutilleul1999mle,manceur2013maximum}. 

The standard EM algorithm has several noticeable issues in our problem of interest.   
In the E-step, we use all the elements in $\mbX_i$ to calculate $\widetilde{\xi}^{(t)}_{ik}$. Even for a tensor of dimension $p_m=10$, $m=1,2,3$, we have one thousand variables and are thus dealing with a high-dimensional estimation problem. 
Even when $Y_i$'s are all observed, we can do no better than random guessing if we estimate an excessive number of parameters without variable selection \citep{bickel2004some,fan2008high}, because the accumulated estimation errors would dominate the signal in the data. Now that $Y_i$'s are unobservable, variable selection should be more critical in order to reduce the number of parameters. Since the standard EM algorithm unfortunately does not enforce variable selection, it is prone to inaccurate clustering on tensor data, which are often high-dimensional (i.e.~$p=\prod_{m=1}^Mp_m > n$). 

In the M-step, an iterative sub-algorithm is needed to maximize $Q_n(\boltheta\mid\widetilde{\boltheta}^{(t)})$ over $\bolSigma_1,\dots,\bolSigma_M$. This sub-algorithm drastically adds to the computation cost. Moreover, the consistency for $\widetilde{\bolSigma}_m^{(t+1)}$ cannot be easily established in high dimensions. To the best of our knowledge, the most related result is \cite{lyu2019tensor}. They considered the estimation of $(\bolSigma_m^*)^{-1}$ under the tensor graphical model where all observations come from the same tensor normal distribution. They had to assume that $(\bolSigma_m^*)^{-1}$ are all sparse and constructed penalized estimates to achieve consistency in high dimensions. The sparsity assumption on the precision matrix is central to their proof. However, in the context of clustering, our goal is to recover $Y_i$'s. The covariances are nuisance parameters for this purpose, as the optimal clustering rule in \eqref{eq.opt} does not depend on $\bolSigma^*_m$ when we know $\mbB^*$. Therefore, it is generally more desirable to not impose additional assumptions on $\bolSigma^*_m$ so that we can handle arbitrary correlation structure while achieving the optimal clustering error. However, the consistency is very difficult to show for the unpenalized estimate $\widetilde{\bolSigma}_m^{(t+1)}$. We need innovative modifications to the M-step to lower the computation cost and achieve theoretical guarantee.

Motivated by the above issues, we propose the doubly-enhanced EM algorithm (DEEM) that greatly improves both the E-step and the M-step in the standard EM algorithm. DEEM consists of iterations between an enhanced E-step and an enhanced M-step. In the enhanced E-step, we impose variable selection to evaluate the Q-function more accurately, while in the enhanced M-step we find better estimates for the covariances. We discuss these two steps in Sections~\ref{Sec:EE} \& \ref{Sec: estmstep}, respectively. The complete DEEM algorithm is summarized in Section~\ref{Sec:DEEM.sub}.  Later in our simulation studies in Section~\ref{Sec: sim}, we confirm that the standard EM algorithm has inferior performance to DEEM.

\subsection{The enhanced E-step}\label{Sec:EE}

To distinguish from the standard EM estimates $\widetilde{\boltheta}$, we denote $\widehat{\boltheta}^{(t)}$ as the DEEM estimate of $\boltheta$ at the $t$-th iteration. Given $\widehat{\boltheta}^{(t)}$, we consider the $(t+1)$-th iteration.

In the enhanced E-step, we obtain a more accurate evaluation of the Q-function in \eqref{eq.Q.function}. 
Obviously, it suffices to estimate $\xi_{ik}^{(t+1)}=\text{Pr}(Y_i=k\mid \mbX_i,\widehat{\boltheta}^{(t)})$. As discussed in Section~\ref{Sec:EM.issues}, estimates of $\xi_{ik}^{(t+1)}$ could contain large estimation error without variable selection. To resolve this issue, we assume that our target $\xi_{ik}=\Pr(Y_i=k\mid \mbX_i,\boltheta^*)$ is determined by a subset of elements in $\mbX$ and hence can be evaluated with a reduced number of parameters. Let 
\beq
\mbB_k^*=\llbracket \bolmu^*_k-\bolmu^*_1;(\bolSigma_1^*)^{-1},\ldots,(\bolSigma_M^*)^{-1}\rrbracket\in\mbbR^{p_1\times\cdots\times p_M}, k=2,\ldots,K.
\eeq
 The following lemma helps clarify the implication of the sparsity assumption. 

\begin{lemma}\label{lem:gammaratio}
Suppose that $\mbX_i$ follows the TNMM \eqref{margdis}. We have that
\bea
\xi_{i1}&=&\dfrac{\pi_1^*}{\pi_1^*+\sum_{k=2}^K \pi_k^*\cdot \exp[\langle\mbX_i-\frac{1}{2}(\bolmu^*_k+\bolmu^*_1),\mbB_k^*\rangle]},\\
\xi_{ik}&=&\dfrac{\pi_k^*\exp[\langle\mbX_i-\frac{1}{2}(\bolmu^*_k+\bolmu^*_1),\mbB_k^*\rangle]}{\pi_1^*+\sum_{k=2}^K \pi_k^*\cdot \exp[\langle\mbX_i-\frac{1}{2}(\bolmu^*_k+\bolmu^*_1),\mbB^*_k\rangle]}, \quad k>1.
\eea
\end{lemma}

Lemma~\ref{lem:gammaratio} shows that each $\xi_{ik}$ is determined by the inner products $\langle\mbX_i-\frac{1}{2}(\bolmu_j^*+\bolmu_1^*),\mbB^*_j\rangle, j=2,\ldots, K$. Hence, $X_{{\cal J}}$ is not important for the E-step
if and only if 
\beq\label{eq.groupsparsity}
 b_{2,{\cal J}}^*=\cdots=b_{K,{\cal J}}^*=0.
\eeq 
Then the sparsity assumption implies that \eqref{eq.groupsparsity} holds for most ${\cal J}$. In other words, let 
${\cal D}$ denote the index set of the important variables, i.e.~$\mathcal{D}^c = \{{\cal J}:  b_{2,{\cal J}}^*=\cdots=b_{K,{\cal J}}^*=0\}$. The sparsity assumption states that $|{\cal D}|\ll \prod_{m=1}^M p_m$. It is worth noting that this assumption is equivalent to assuming that the optimal clustering rule is sparse. By \eqref{eq.opt.generic}, the optimal rule under TNMM is
\beq\label{eq.opt.TNMM}
\widehat{Y}^{opt}=\arg\max_k\{\log\pi_k^*+\langle\mbX-(\bolmu_1^*+\bolmu_k^*)/2,\mbB_k^*\rangle\},
\eeq
where $\mbB_1^*=0$.
Hence, the sparsity in the optimal rule concurs with our assumption on $\calD$, where variable selection assists in achieving the lowest clustering error possible.

Note that $\mbB_k^*=\llbracket\bolmu_k^*-\bolmu_1^*;(\bolSigma^*_1)^{-1},\ldots,(\bolSigma_M^*)^{-1}\rrbracket$ by definition. It follows that
\begin{equation}\label{bori}
(\mbB_2^*,\ldots,\mbB^*_K)=\argmin_{\mbB_2,\dots,\mbB_K\in\mathbb{R}^{p_1\times\cdots p_M}}\bigg[\sum_{k=2}^K(\langle\mbB_k,\llbracket \mbB_k,\bolSigma^*_1,\ldots,\bolSigma^*_M\rrbracket\rangle-2\langle\mbB_k,\bolmu_k^*-\bolmu^*_1\rangle)\bigg].
\end{equation}

To obtain sparse estimates for $\mbB_k^*$, we plug in our current estimates for $\bolSigma^*_m$ and $\bolmu^*_k$, and add the group lasso penalty \citep{yuan2006model} to encourage the sparsity pattern in \eqref{eq.groupsparsity}. More specifically, we let $(\hatmbB_2^{(t+1)},\ldots,\hatmbB_K^{(t+1)})$ be the solution to the following minimization problem,
\begin{equation}\label{betaopt}
\min_{\mbB_2,\dots,\mbB_K}\bigg[\sum_{k=2}^K(\langle\mbB_k,\llbracket \mbB_k,\hatbolSigma_1^{(t)},\ldots,\hatbolSigma_M^{(t)}\rrbracket\rangle-2\langle\mbB_k,\hatbolmu_k^{(t)}-\hatbolmu_1^{(t)}\rangle) +\lambda^{(t+1)}\sum_{{\cal J}}\sqrt{\sum_{k=2}^Kb_{k,{\cal J}}^2}\bigg],
\end{equation}
where $\lambda^{(t+1)}>0$ is a tuning parameter. The optimization problem in \eqref{betaopt} is convex and can be easily solved by a blockwise coordinate descent algorithm similar to that in \citet{CATCH}. See Algorithm~S.2 in Section~B in Supplementary Materials for details.


After obtaining $\{\hatmbB^{(t+1)}_2,\ldots,\hatmbB^{(t+1)}_K\}$, we calculate
\bea
\widehat\xi_{i1}^{(t+1)}&=&\dfrac{\widehat\pi^{(t)}_1}{\widehat\pi^{(t)}_1+\sum_{k=2}^K \widehat\pi^{(t)}_k\cdot \exp[\langle\mbX_i-\frac{1}{2}(\widehat\bolmu^{(t)}_k+\widehat\bolmu^{(t)}_1),\widehat\mbB^{(t+1)}_k\rangle]},\label{eq.xi1hat}\\
\widehat\xi_{ik}^{(t+1)}&=&\dfrac{\widehat\pi_k^{(t)}\exp[\langle\mbX_i-\frac{1}{2}(\widehat\bolmu_k^{(t)}+\widehat\bolmu^{(t)}_1),\widehat\mbB^{(t+1)}_k\rangle]}{\widehat\pi^{(t)}_1+\sum_{k=2}^K \widehat\pi^{(t)}_k\cdot \exp[\langle\mbX_i-\frac{1}{2}(\widehat\bolmu^{(t)}_k+\widehat\bolmu^{(t)}_1),\widehat\mbB^{(t+1)}_k\rangle]},\quad k>1\label{eq.xikhat}.
\eea

Combining $\widehat\xi_{ik}^{(t+1)}$ with \eqref{classdensity} and \eqref{eq.Q.function}, we have the Q-function in the $(t+1)$-th iteration as 
	\begin{equation}\label{Qfunctensor}
	Q^{\text{DEEM}}(\boltheta\mid\widehat{\boltheta}^{(t)})=\sum_{i=1}^n\sum_{k=1}^K\widehat\xi_{ik}^{(t+1)}\{\log\pi_k-(\sum_{m=1}^M q_m\log\lvert \bolSigma_m\rvert)-\frac{1}{2}\langle \llbracket \mbX_i-\bolmu_k;\bolSigma_1^{-1},\ldots,\bolSigma_M^{-1}\rrbracket, \mbX_i-\bolmu_k\rangle\}.
	\end{equation}
The Q-function in \eqref{Qfunctensor} will guide us to find $\widehat{\boltheta}^{(t+1)}$ in the enhanced M-step, which will be discussed in Section~\ref{Sec: estmstep}.
Since the probabilities $\widehat{\xi}_{ik}^{(t+1)}$ in \eqref{Qfunctensor} are calculated based on a small subset of variables, they are expected to be close to the truth under the sparsity model assumption, and lay the foundation for accurate parameter estimation in the enhanced M-step.



\subsection{The enhanced M-step}\label{Sec: estmstep}
In the enhanced M-step, we update estimates for $\pi_k^*,\bolmu_k^*$ and $\bolSigma_m^*$. By maximizing the Q-function in \eqref{Qfunctensor}, it is straightforward to obtain the estimates for $\pi_k^*,\bolmu_k^*$ at the $(t+1)$-th iteration as $\hatpi_k^{(t+1)}={\sum_{i=1}^n\widehat{\xi}_{ik}^{(t+1)}}/{n}$ and $\hatbolmu_k^{(t+1)}={\sum_{i=1}^n\widehat{\xi}_{ik}^{(t+1)}\mbX_i}/{\sum_{i=1}^n\widehat{\xi}_{ik}^{(t+1)}}$, $k=1,\dots,K$. It is also easy to verify that $\sum_{k=1}^K\widehat\pi_k^{(t+1)}=\sum_{k=1}^K\widehat{\xi}_{ik}^{(t+1)}=1$.

%
%
%
As discussed in Section~\ref{Sec:EM.issues}, directly maximizing the Q-function over $\bolSigma_m$ is not ideal. We consider an alternative update for ${\bolSigma}_m$ based on the following result. Recall that $q_m=p_m^{-1}\prod_{h=1}^M p_h$ and $\xi_{ik}=\Pr(Y_i=k\mid \mbX_i,\boltheta^*)$.  
\begin{lemma}\label{lem.sigmam.expectation}
Under the TNMM in \eqref{margdis}, we have
\beq\label{eq:sigmam.expectation}
\bolSigma_m^*\propto \frac{1}{q_m}\E\left\{\sum_{k=1}^K\xi_{ik}(\mbX_i-\bolmu_k^*)_{(m)}(\mbX_i-\bolmu_k^*)_{(m)}\T\right\}.
\eeq
\end{lemma}
Lemma~\ref{lem.sigmam.expectation} implies that we can construct a method of moment estimate for $\bolSigma_m^*$. Recall that we require $\sigma^*_{m,11}=1$ for $m>1$ to ensure identifiable covariance matrices and hence have $\sigma_{1,11}^*=\Var(X_{i,1\cdots 1}\mid Y_i=k)$, where $X_{i,1\cdots 1}$ is the $(1,\ldots,1)$-th element of $\mbX_i$. Since we have shown that $\bolSigma_m^*$ is proportional to the right hand side of \eqref{eq:sigmam.expectation}, 
we can incorporate the identification constraints into scaling. Note that Lemma~\ref{lem.sigmam.expectation} is widely applicable and can be combined with any other identification constraints on $\bolSigma_m^*$, e.g.~requiring $\Vert\bolSigma_2^*\Vert_F=\dots=\Vert\bolSigma_M^*\Vert_F=1$. 

As a consequence of Lemma~\ref{lem.sigmam.expectation}, we propose the following non-iterative estimator for the covariance parameters in the enhanced M-step. Given $\widehat{\xi}_{ik}^{(t+1)}$, we first compute intermediate estimates,
\beq
\widebreve{\bolSigma}_{m}^{(t+1)}=\frac{1}{nq_m}\sum_{i=1}^n\sum_{k=1}^K\widehat\xi^{(t+1)}_{ik}(\mbX_i-\widehat\bolmu_k^{(t+1)})_{(m)}(\mbX_i-\widehat\bolmu^{(t+1)}_k)_{(m)}\T, \quad m=1,\dots,M.
\eeq
Then, for $m>1$, our DEEM estimator is $\widehat{\bolSigma}^{(t+1)}_m=\widebreve{\bolSigma}^{(t+1)}_m/\widebreve{\sigma}^{(t+1)}_{m,11}$; and for $m=1$, our DEEM estimator is $\widehat{\bolSigma}^{(t+1)}_m=\{\widehat\sigma^{(t+1)}_{1,11}/\widebreve{\sigma}^{(t+1)}_{1,11}\} \widebreve{\bolSigma}^{(t+1)}_1$, where the conditional variance of the element $X_{i,1\cdots 1}$ is estimated as
\beq
\widehat\sigma^{(t+1)}_{1,11}=\widehat{\E}\{\widehat{\Var}(X_{i,1\cdots 1}\mid Y_i)\}=\frac{1}{n}\sum_{i=1}^n\sum_{k=1}^K\widehat\xi_{ik}^{(t+1)}(X_{i,{1 \cdots 1}}-\widehat\mu^{(t+1)}_{k,{1 \cdots 1}})^2.
\eeq

The covariance estimates $\widehat{\bolSigma}_m^{(t+1)}$ will be used in the subsequent enhanced E-step. In comparison to the estimator $\widetilde{\bolSigma}_m^{(t+1)}$ in the standard EM algorithm, $\widehat{\bolSigma}_m^{(t+1)}$ has apparent computational advantages. No iterative sub-algorithm is needed for computing $\widehat{\bolSigma}_m^{(t+1)}$. Instead, all the computation in the enhanced M-step can be carried out explicitly. Moreover, we will later show that $\widehat{\bolSigma}_m^{(t+1)}$ leads to consistent clustering even when the dimension of each mode of the tensor grows at an exponential rate of $n$ without any sparsity assumption on $\bolSigma_m^*$. It is unclear whether such consistency can be achieved by the standard estimator $\widetilde{\bolSigma}_m^{(t+1)}$. Hence, $\widehat{\bolSigma}_m^{(t+1)}$ should be preferred over $\widetilde{\bolSigma}_m^{(t+1)}$ for theoretical considerations as well.

The enhanced M-step in DEEM is delicately designed, but our covariance estimator has a potentially much wider range of applications beyond DEEM. For example, \citet{Cao2013,Sun2018} considered combining low-rank decomposition of the tensors and the K-means clustering; \citet{Gao2018} proposed to regularize the mean differences of matrix observations. To fill the gap between these works and the optimal clustering rule, which requires covariance modeling, one can potentially adopt our fast and theoretically guaranteed covariance estimators.



\subsection{The DEEM algorithm and implementation details}\label{Sec:DEEM.sub}

\begin{algorithm}[!t]
	\begin{enumerate}
		\item Initialize $\hatpi_k^{(0)}$, $\hatbolmu_k^{(0)}$, $\hatbolSigma_m^{(0)}$.
		\item For $t=0,1,\ldots$, repeat the following steps until convergence:
		\begin{enumerate}
			\item The enhanced E-step:
			\begin{enumerate}
				\item Minimize the following convex objective function over ${\mbB_2,\dots,\mbB_K\in\mbbR^{p_1\times\cdots\times p_M}}$ with Algorithm~S.2:		
				$$\sum_{k=2}^K(\langle\mbB_k,\llbracket \mbB_k,\hatbolSigma_1^{(t)},\ldots,\hatbolSigma_M^{(t)}\rrbracket\rangle-2\langle\mbB_k,\hatbolmu_k^{(t)}-\hatbolmu_1^{(t)}\rangle) +\lambda^{(t+1)}\sum_{{\cal J}}\sqrt{\sum_{k=2}^Kb_{k,{\cal J}}^2}.$$				
				Let $(\hatmbB^{(t+1)}_2,\cdots,\hatmbB_K^{(t+1)})$ denote the solution.		
				\item For $i=1,\dots,n$, and $k=1,\dots,K$, calculate the probabilities
				
				$$
				\widehat\xi_{ik}^{(t+1)} = \begin{cases}
		\dfrac{\widehat\pi^{(t+1)}_1}{\widehat\pi^{(t+1)}_1+\sum_{j=2}^K \widehat\pi^{(t+1)}_j\cdot \exp[\langle\mbX_i-\frac{1}{2}(\widehat\bolmu^{(t+1)}_j+\widehat\bolmu^{(t+1)}_1),\hatmbB^{(t+1)}_j\rangle]},& k=1;\\
		\dfrac{\widehat\pi_k^{(t+1)}\exp[\langle\mbX_i-\frac{1}{2}(\widehat\bolmu_k^{(t+1)}+\widehat\bolmu^{(t+1)}_1),\hatmbB^{(t+1)}_k\rangle]}{\widehat\pi^{(t+1)}_1+\sum_{k=2}^K \widehat\pi^{(t+1)}_j\cdot \exp[\langle\mbX_i-\frac{1}{2}(\widehat\bolmu^{(t+1)}_j+\widehat\bolmu^{(t+1)}_1),\hatmbB^{(t+1)}_j\rangle]},              & k>1.
		\end{cases}
				$$
%
			\end{enumerate}
		\item The enhanced M-step:
		\begin{enumerate}
		\item Update $\hatpi_k^{(t+1)}={\sum_{i=1}^n\widehat{\xi}_{ik}^{(t+1)}}/{n}$ and $\hatbolmu_k^{(t+1)}={\sum_{i=1}^n\widehat{\xi}_{ik}^{(t+1)}\mbX_i}/{\sum_{i=1}^n\widehat{\xi}_{ik}^{(t+1)}}$.
			
			\item Compute intermediate covariance estimators \beqn
\widebreve\bolSigma_m^{(t+1)}=\frac{1}{nq_m}\sum_{i=1}^n\sum_{k=1}^K\widehat\xi^{(t+1)}_{ik}(\mbX_i-\widehat\bolmu_k^{(t+1)})_{(m)}(\mbX_i-\widehat\bolmu^{(t+1)}_k)_{(m)}\T.
\eeqn	
\item Scale $\widebreve\bolSigma_m^{(t+1)}$ to be
$$
				\widehat{\bolSigma}^{(t+1)}_m = \begin{cases}
		\{n^{-1}\sum_{i=1}^n\sum_{k=1}^K\widehat\xi_{ik}^{(t+1)}(X_{i,{1 \cdots 1}}-\widehat\mu^{(t+1)}_{k,{1 \cdots 1}})^2\}\widebreve{\bolSigma}^{(t+1)}_m /\widebreve{\sigma}^{(t+1)}_{1,11},& m=1;\\
		\widebreve{\bolSigma}^{(t+1)}_m/\widebreve{\sigma}^{(t+1)}_{m,11},              & m>1.
		\end{cases}
				$$

%
%
%
		\end{enumerate}
	\end{enumerate}
	\item Output $\widehat{\xi}_{ik},\widehat{\pi}_k,\widehat{\bolmu}_k,\widehat{\bolSigma}_m$ at convergence.
	\end{enumerate}
	\caption{DEEM algorithm for tensor clustering}
	\label{Algorithm tensor clustering}
\end{algorithm}

With the enhanced E-step and the enhanced M-step, we iterate between them until convergence similar to the standard EM algorithm. The DEEM algorithm is summarized in Algorithm~\ref{Algorithm tensor clustering}. Given the output of $\widehat{\xi}_{ik}$, we assign $\mbX_i$ to cluster $\widehat Y_i^{\text{DEEM}}$, where $\widehat Y_i^{\text{DEEM}}=\arg\max_{k} \widehat{\xi}_{ik}$. We further discuss some implementation details in what follows.


\textbf{Initialization. }In order to implement DEEM, we need to determine the initial value. In our numerical studies, we first perform the K-means clustering on $\vecc{(\mbX_i)}$ to find $\widehat Y^{(0)}_i$. Then we set the initial values as,
\bea
&&\widehat{\pi}_k^{(0)}=\frac{\sum_{i=1}^n \mathrm{1}(\widehat Y^{(0)}_i=k)}{n},\quad\widehat{\bolmu}_k^{(0)}=\frac{1}{n\widehat{\pi}_k^{(0)}}\sum_{\widehat Y^{(0)}_i=k} \mbX_i,\\
&&\widehat{\bolSigma}^{(0)}_m\propto\frac{1}{nq_m}\sum_{i=1}^n\sum_{\widehat Y^{(0)}_i=k}(\mbX_i-\widehat{\bolmu}_k^{(0)})_{(m)}(\mbX_i-\widehat{\bolmu}_k^{(0)})_{(m)}\T.
\eea
The scales of $\widehat{\bolSigma}^{(0)}_m$ are chosen such that $\widehat{\sigma}^{(0)}_{m,11}=1$ for $m>1$ and $\widehat{\sigma}^{(0)}_{1,11}=\frac{1}{n}\sum_{k=1}^K\sum_{\widehat Y_i^{(0)}=k}(X_{i,1\cdots 1}-\widehat{\mu}_k^{(0)})^2$. We choose to use the K-means clustering in initialization because it is very fast. {Although the K-means clustering is performed on vectorized data and thus ignores the tensor structure, DEEM can recover from this loss of efficiency by incorporating the tensor structure in the later iterations. }In our numerical studies, we observe that this initialization leads to good solutions of DEEM at convergence even when the K-means clustering has poor performance. 

\textbf{Convergence.} In our implementation, the convergence criterion is based on the sum of squares of mean differences between two consecutive iterations. We stop the DEEM iterations  if $\sum_k\Vert\hatbolmu_k^{(t+1)}-\hatbolmu_k^{(t)} \Vert_F^2\le 0.1$ or the maximum number of iterations $t_{\max} = 50$ is reached. In our experience, the algorithm usually converges within $50$ iterations. See Section~\ref{Sec: Delta}, Figure~\ref{fig:M1_delta}, for the number of iterations required to converge as we change the signal strength in simulations.

\textbf{Tuning.} The tuning parameter $\lambda^{(t)}$ in the enhanced E-step could either be fixed or varying across iterations. For computation considerations, it is apparently easier to fix $\lambda^{(t)}=\lambda$ for all $t$, while for theoretical considerations, one may favor varying $\lambda^{(t)}$. For example, similar to \citet{Cai2017CHIME}, we could consider 
\beq\label{eq.lambda.t}
\lambda^{(t+1)}=\kappa\lambda^{(t)}+(\frac{1-\kappa^{t+1}}{1-\kappa})C_{\lambda}\sqrt{\frac{\log{p}}{n}},
\eeq
 where $0<\kappa<1/2$ and $C_{\lambda}>0$ are constants. In our numerical studies, we note that both choices of $\lambda^{(t)}$ give reasonable results as long as they are properly tuned. Moreover, when $t$ is large, the varying $\lambda^{(t)}$ in \eqref{eq.lambda.t} is roughly constant at the value $C_{\lambda}\sqrt{\frac{\log{p}}{n}}$. Therefore, we fix $\lambda^{(t)}=\lambda$ in all the numerical studies, but only consider the varying $\lambda^{(t)}$ in theoretical studies.

 When we fix $\lambda^{(t)}=\lambda$, we need to determine $\lambda$. Permutation \citep[e.g.,][]{Witten2010SKM} and Bayesian information criterion \citep[BIC; e.g.,][]{Sun2018, Guo2010} are two popular ways for tuning in clustering. We adopt a BIC-type criterion. For any $\lambda$, we let $\widehat{\boltheta}^{\lambda}$ be the output of DEEM with the tuning parameter fixed at $\lambda$. We look for the value of $\lambda$ that minimizes
\begin{equation}\label{bic}
\text{BIC}(\lambda)=-2\sum_{i=1}^n\log(\sum_{k=1}^K\widehat{\pi}^{\lambda}_kf_k(\mbX_i;\hatboltheta_k^{\lambda}))+\log(n)\cdot \lvert \widehat{\mathcal{D}}^{\lambda}\rvert,
\end{equation}
where $\widehat{\mathcal{D}}^{\lambda}=\{(k, {\cal J}): \widehat b_{k,{\cal J}}^{\lambda} \neq 0 \}$ is the set of nonzero elements in $\widehat{\mbB}_2^{\lambda},\ldots,\widehat{\mbB}_K^{\lambda}$.

\textbf{Number of clusters.}  As most clustering methods, DEEM requires users to specify the number of clusters $K$, the knowledge of which is often unavailable due to the unsupervised nature of clustering problems. In this paper, we focus on the scenario that $K$ is known. In practice, we may use a  BIC-type criterion similar to \eqref{bic} to choose $\lambda$ and $K$ simultaneously. Implementation details and simulation examples of this approach are provided in Section~C of Supplementary Materials. Under simulation models M1--M5 in Section~\ref{Sec: num}, the number of clusters can be identified correctly for roughly 60\% to 80\% of the time.  There exist many proposals for the estimation of $K$ in various clustering contexts, such as \citet{tibshirani2001estimating, review1,sugar2003finding,chiang2010intelligent, wang2010consistent,fang2012selection,fujita2014non,fu2020estimating}, but consistent selection of $K$ for high-dimensional tensor clustering is still an open question and is left as future research.

 As pointed out by a referee, it has been shown in more classical settings that if $K$ is over-specified, the convergence rate could be lower \citep[e.g]{chen1995optimal,heinrich2018strong,dwivedi2018singularity}. These papers focus on vector models and consider dimensions much lower than what will be presented for our method. It will be an interesting but challenging future topic to know whether similar results hold for tensor clustering in high dimensions.
 

\subsection{Benefits of keeping the tensor form}\label{Sec: benefit.tensor}

In this section, we discuss the advantages of keeping the tensor form in developing our method. We consider the enhanced E-step and the enhanced M-step separately.

Our enhanced E-step is conceptually similar to the E-step in \citet{Cai2017CHIME}, where they proposed a method called CHIME for model-based clustering of high-dimensional vector data. Under the high-dimensional GMM, the authors showed that it suffices to find some linear projections of the features to conduct the E-step. To tackle the high dimensionality, they assume that the linear projections are sparse. Their sparsity assumption is similar to ours on $\mbB_k^*$. 

However, CHIME is only designed for vector data, and is not tailored for tensor data. Our enhanced E-step takes advantage of the tensor structure to reduce the storage cost and improve clustering efficiency for higher-order tensor data. In particular, if we vectorize our tensor observations $\mbX_i$ and apply CHIME, in each iteration we need to compute the covariance matrix $\widehat{\Var}(\vecc(\mbX_i)\mid Y_i)$ with $\prod_{m=1}^M p_m^2$ elements. But in our enhanced E-step, the covariance matrices only have $\sum_{m=1}^M p_m^2$ elements, and are much  lighter on the storage. 

On the other hand, even though the vectorized form of TNMM in \eqref{eq.TNMM.vec} has a reduced number of parameters, it is still advantageous to consider the original tensor form for the sake of computation. To see this subtle point, note that, if we vectorize $\mbX_i$ and the associated parameters $\bolbeta_k^*\equiv \vecc(\mbB^*)=\left\{\bigotimes_{m=M}^{m=1}(\bolSigma_m^*)^{-1}\right\}\vecc(\bolmu_k^*-\bolmu_1^*)$. Then the optimization problem \eqref{betaopt} becomes
\begin{equation}\label{eq.betaopt.vec}
\argmin_{\bolbeta_2,\ldots,\bolbeta_K\in\mbbR^{p}}\left[\sum_{k=2}^K\left\{\bolbeta_k\T\left(\bigotimes_{m=M}^{m=1}\widehat{\bolSigma}_{m}^{(t)}\right)\bolbeta_k-2\bolbeta_k\T\vecc\left(\widehat\bolmu_k^{(t)}-\widehat\bolmu_1^{(t)}\right)\right\} +\lambda^{(t+1)}\sum_{j}\sqrt{\sum_{k=2}^K\beta_{k,j}^2}\right],
\end{equation}
which can be solved by a blockwise coordinate descent algorithm, such as the one in \citet{msda}. However, the storage and the computation costs of $\bigotimes_{m=M}^{m=1}\widehat{\bolSigma}_{m}^{(t)}$ are both at the intimidating order of $O(\prod_{m=1}^Mp_m^2)$. In comparison, to solve \eqref{betaopt}, our efficient implementation does not require calculating the Kronecker product. Consequently, it may be practically infeasible to solve \eqref{eq.betaopt.vec} when we can still easily solve \eqref{betaopt}. For example, on a simulated data set from M7 in Section~\ref{Sec: sim}, we tried to use \eqref{eq.betaopt.vec} in the enhanced E-step when the tensor dimension is $30\times 30\times 30$. On a computer within 16GB of memory, the algorithm would fail due to an out-of-memory error. However, DEEM can be carried out on the same computer.

More importantly, the enhanced M-step is most naturally derived when we keep the tensor form of $\mbX_i$ rather than considering the vectorized TNMM in \eqref{eq.TNMM.vec}. With the tensor form, the covariance matrices $\bolSigma_m^*$ are more ``separated'' from each other. This fact enables us to find $\widehat{\bolSigma}_m^{(t)}$ individually. If we consider the vectorized version, we need to estimate $\bigotimes_{m=M}^{m=1}\bolSigma_m^*$, which is not easy without reshaping $\vecc(\mbX_i)$ into tensors.

\section{Theoretical studies}\label{Sec: theory}
\subsection{Parameter space and technical definitions}

Before presenting the consistency of DEEM, we define our parameter space of interest and formally introduce some technical terms. We assume that the number of cluster is known and focus on the two-cluster case, i.e.~$K=2$.

We define our parameter space for the TNMM parameter as $
\boltheta=\{\pi_1,\pi_2,\bolmu_1,\bolmu_2,\bolSigma_1,\dots,\bolSigma_M\}$,
where $0<\pi_1=1-\pi_2<1$, $\bolmu_1, \bolmu_2\in\mathbb{R}^{p_1\times\cdots\times p_M}$ and $\bolSigma_1,\dots,\bolSigma_M\in\mathbb{R}^{p_m\times p_m}$ are all symmetric positive definite. Two important estimable functions of $\boltheta$ are $\mbB=\mbB(\boltheta)=\llbracket\bolmu_2-\bolmu_1;\bolSigma_1^{-1},\ldots,\bolSigma_M^{-1}\rrbracket\in\mbbR^{p_1\times\cdots\times p_M}$ and $\Delta=\Delta(\boltheta)=\langle\bolmu_2-\bolmu_1,\llbracket\bolmu_2-\bolmu_1;\bolSigma_1^{-1},\ldots,\bolSigma_M^{-1}\rrbracket\rangle\in\mbbR$. The tensor parameter $\mbB$ is used for calculating $\widehat\xi_{ik}$ in the enhanced E-step of DEEM. The parameter $\Delta$ is the separation between the two clusters. The true population parameters are $\boltheta^*$, $\mbB^*=\mbB(\boltheta^*)$ and  $\Delta^*=\Delta(\boltheta^*)$. The set of important variables $\calD=\{\mathcal{J}: b_{\cal J}\neq0 \}$ is clearly also an estimable function of $\boltheta$, as $\calD=\calD(\mbB)=\calD(\mbB(\boltheta))$.

For any two numbers $a$ and $b$, we write $a\lor b=\max\{a,b\}$ and $a\land b=\min\{a,b\}$. We use $\lambda_{\max}(\cdot)$ and  $\lambda_{\min}(\cdot)$ to denote the largest and the smallest eigenvalues of a matrix, respectively. We define the parameter space $\bolTheta=\bolTheta(c_{\pi},C_b,s,\{C_m\}_{m=1}^{M=1},\Delta_0)$ as
\begin{equation}\label{eq.param}
\{\boltheta:  \pi_k\in(c_{\pi},1-c_{\pi}), \Vert\vecc(\mbB)\Vert_1\le C_b, \lambda_{\min}^{-1}(\bolSigma_m)\lor\lambda_{\max}(\bolSigma_m)\le C_m, \vert\calD\vert \le s, \Delta\ge\Delta_0
\},
\end{equation}
where $C_1,\ldots, C_M, C_b, \Delta_0>0$ and $0<c_\pi<1$ are constants that do not change as $p_m$ increases, but $s>0$ can vary with $p_m$. This parameter space is sufficiently flexible to include a wide range of models. The assumptions in $\bolTheta$ are intuitive and very mild. First, we require the eigenvalues of $\bolSigma_m$ to be bounded from below and above: $C_m^{-1}\le\lambda_{\min}(\bolSigma_m) \le\lambda_{\max}(\bolSigma_m)\le C_m$. This eigenvalue assumption on the covariances is also common in high dimensions \citep{CL2011,CATCH}. We also require that $\pi_k$ is bounded away from $0$ and $1$, so that each cluster has a decent sample size. The coefficient $\mbB$ is assumed to be sparse so that we can perform variable selection. Finally, the assumption that $\Delta>\Delta_0$ implies that the two clusters are well separated from each other. If two clusters are indistinguishable even on the population level, of course it will be impossible to separate them with any clustering rule. In our theory, we need $\Delta_0$ to be sufficiently large so that we only consider models with a reasonably large separation. See more detailed discussion of the cluster separation $\Delta$ following Theorem~\ref{thm.B}.

We need some more technical definitions before we present the conditions needed for theoretical results. We set 
\beq\label{eq.Gamma}
\Gamma(s)=\{\mbu\in\mathbb{R}^p: 2\Vert\mbu_{S^C}\Vert_1\le 4\Vert\mbu_{S}\Vert_1+3\sqrt{s}\Vert\mbu\Vert_2, \mbox{ for some $S\subset \{1,\ldots,p\}$, $|S|=s$}\},
\eeq
where $\mbu_{S}\in\mbbR^s$ and $\mbu_{S^C}\in\mbbR^{p-s}$ are sub-vectors extracted from $\mbu$ based on the index set $S$ and its complement set.
The set $\Gamma(s)$ contains approximately sparse vectors with at most $s$ elements well separated from 0.
For a vector $\mba\in\mathbb{R}^{p}$ and a matrix $\mbA\in\mathbb{R}^{p\times p}$, we denote
\beq
\Vert\mba\Vert_{2,s}=\sup_{\Vert\mbx\Vert_2=1,\mbx\in \Gamma(s)}|\mba\T\mbx|, \quad \Vert\mbA\Vert_{2,s}=\sup_{\Vert\mbx\Vert_2=1,\mbx\in \Gamma(s)}\Vert\mbA\mbx\Vert_2.
\eeq 

For two parameters $\boltheta$ and $\widetilde\boltheta$, we define their distance as:
\beq\label{eq.d2s}
d_{2,s}(\boltheta,\widetilde\boltheta)=(\lor_k|\pi_k-\widetilde{\pi}_k|)\lor(\lor_k\Vert\vecc(\bolmu_k-\widetilde\bolmu_k)\Vert_{2,s})\lor\Vert(\otimes_{m=M}^{m=1}\bolSigma_m-\otimes_{m=M}^{m=1}\widetilde{\bolSigma}_m)\vecc(\widetilde{\mbB})\Vert_{2,s},
\eeq 
where $\lor_k a_k = \max\{ a_k: k=1,2,\dots\}$.

We further define the contraction basin for $\boltheta^*$ as
\bea
&&{\cal B}_{con}(\boltheta^*; a_{\pi},a_{\Delta},a_b,s) \label{eq.B.contr} \\
&=&\{\boltheta:
\pi_k\in (a_{\pi},1-a_{\pi}), (1-a_{\Delta})(\Delta^*)^2<|\delta_k(\mbB)|,\sigma^2(\mbB)<(1+a_{\Delta})(\Delta^*)^2,\nonumber\\
&&\vecc(\mbB-\mbB^*)\in\Gamma(s),\Vert\vecc(\mbB-\mbB^*)\Vert_1\le a_b\Delta^*,\Vert\vecc(\bolmu_k)\Vert_{2,s}\le a_b\Delta^*\},\nonumber
\eea
where $a_{\pi},a_{\Delta},a_b>0$, $a_{\pi}\le c_{\pi}<1$ are constants, $\delta_k(\mbB)=\langle\mbB,\bolmu_k^*-({\bolmu_1+\bolmu_2})/{2}\rangle$ and $\sigma^2(\mbB)=\langle\mbB,\llbracket\mbB,\bolSigma_1^*,\ldots,\bolSigma_M^*\rrbracket\rangle$.

\subsection{Initialization condition}\label{Sec: C1}

We introduce a condition on the initial value that is important for our study. 
Define $d_0=d_{2,s}(\widehat\boltheta^{(0)},{\boltheta}^*)$ as the distance between the initial value $\widehat\boltheta^{(0)}$ and the true parameter ${\boltheta}^*$, where the function $d_{2,s}(\cdot,\cdot)$ is defined in \eqref{eq.d2s}.
For an $M$-way tensor $\mbA\in\mathbb{R}^{p_1\times\cdots\times p_M}$, we let $\Vert\mbA\Vert=\sqrt{\sum_{{\cal J}} A_{\mathcal{J}}^2}$. The consistency of DEEM relies on the following condition, where $a_{\pi}$ and $a_{\Delta}$ are defined in \eqref{eq.B.contr}, $c_{\pi}$ and $C_0\equiv\prod_{m=1}^M C_m$ are from \eqref{eq.param}.
\begin{enumerate}
\item[(C1)] The initial estimator $\widehat{\mbB}^{(0)}$ satisfies that $d_0\lor \Vert\widehat{\mbB}^{(0)}-\mbB^*\Vert\le r\Delta^*,  \vecc(\widehat{\mbB}^{(0)}-\mbB^*)\in\Gamma(s)$, where $r<\min\{ \frac{|a_{\pi}-c_{\pi}|}{\Delta}, \frac{\sqrt{9C_0+16a_{\Delta}}-\sqrt{9C_0}}{4},  \frac{a_{\Delta}}{C_0},  \frac{a_b}{5\sqrt{s}} \}$ and $r^{-1} = o\left(\sqrt{n/s\sum_{m=1}^M\log{p_m}}\right)$.
\end{enumerate}

Condition~(C1) indicates that the initial value is reasonable in the sense that $d_0$ is relatively small, and $\widehat\mbB^{(0)}$ is close to $\mbB^*$ and approximately sparse. This condition is important for our theoretical study because it guarantees that each iteration keeps improving our estimate. Due to the non-convex nature of clustering analysis, conditions on the initial value are popular in its theoretical studies; see \citet{wang2015high,yi2015regularized,balakrishnan2017statistical,Cai2017CHIME} for example. Finding good initial values for cluster analysis is an important research area on its own, with many interesting works for the Gaussian mixture model \citep{kalai2010efficiently,moitra2010settling,hsu2013learning,hardt2015tight}. 

For theoretical interests, we show that there exists an algorithm to generate initial values satisfying Condition~(C1). One such initialization algorithm is presented as Algorithm~S.4 in Section~G of Supplementary Materials. Algorithm~S.4 is related to the vector-based algorithm in \citet{hardt2015tight}, but is specially designed for tensor data. Under TNMM, it produces initial values that satisfy Condition~(C1) under appropriate conditions, as shown in the following lemma.
\begin{lemma}\label{lem.C1}
Under the TNMM in \eqref{margdis}, suppose $\boltheta^*\in\bolTheta(c_{\pi},C_b,s,\{C_m\}_{m=1}^M,C_b,\Delta_0)$. If $s^{12}\sum_{m=1}^M\log{p_m}=o(n)$, with a probability greater than $1-O(\prod_{m}p_m^{-1})$, Algorithm~S.4 produces initial values that satisfy Condition~(C1).
\end{lemma}
Lemma~\ref{lem.C1} indicates that, under TNMM, when the sample size $n$ is larger than $s^{12}\sum_{m=1}^M\log{p_m}$, Condition~(C1) is satisfied by Algorithm~S.4 with a probability tending to 1 as $n\rightarrow \infty$. Hence, we can meet Condition~(C1) even when the dimension of each mode grows at an exponential rate of the sample size. {The term $s^{12}$ results from the theoretical properties of the initialization algorithm proposed by \citet{hardt2015tight}. Their algorithm solves an equation system that involves the first six moments of Gaussian mixtures. We need $s$ to grow at our specified rate such that all these moments are estimated accurately. }Also note that this sample size requirement matches the best one in literature when $M=1$ and tensors reduce to vectors.

In the literature, there are also interests in removing conditions for initial values completely \citep{daskalakis2017ten,wu2019randomly}. All these works require extensive efforts, and there is a considerable gap between these works and  the topic in the manuscript. The existing works focus on low-dimensional vectors with known covariance matrices that are often assumed to be identity matrices, while we have high-dimensional tensors with unknown covariance matrices. 

\subsection{Main theorems}

For our theory, we assume that the tuning parameters in DEEM are generated according to \eqref{eq.lambda.t}, with $\lambda^{(0)}$ defined as
\beq\label{eq.lambda0}
\lambda^{(0)}=C_d\cdot(|\widehat{\pi}_2|\lor \Vert\vecc(\hat\bolmu_1^{(0)}-\widehat{\bolmu}_2^{(0)})\Vert_{2,s}\lor \Vert\bigotimes_{m=M}^{m=1}\widehat{\bolSigma}_{m}^{(0)}\Vert_{2,s})/\sqrt{s}+C_{\lambda}\sqrt{\sum_{m=1}^M\log{p_m}/n},
\eeq
where $C_d,C_{\lambda}>0$ are constants.

Our ultimate goal is to show that the DEEM is asymptotically equivalent to the optimal rule in terms of clustering error. However, because $\mbB^*$ is the key parameter in clustering, we first present the theoretical properties of $\widehat{\mbB}^{(t)}$ as an intermediate result. 

\begin{theorem}\label{thm.B}
Consider $\boltheta^*\in\bolTheta(s,c_{\pi},\{C_m\}_{m=1}^M,C_b,\Delta_0)$ with $s=o(\sqrt{n/\sum_m\log{p_m}})$ and a sufficiently large $\Delta_0$. Assume that Condition (C1) holds with $\sqrt{\sum_m\log{p_m}/n}=o(r)$, $\lambda^{(0)}$ is specified as in \eqref{eq.lambda0} and $\lambda^{(t)}$ is specified as in \eqref{eq.lambda.t}. Then there exist constants $C_d,C_{\lambda}>0$ and $0<\kappa<1/2$ such that, with a probability greater than $1-O(\prod p_m^{-1})$, we have
\beq\label{eq.B.t}
\Vert\widehat{\mbB}^{(t)}-\mbB^*\Vert\lesssim \kappa^td_0+\sqrt{\frac{s\sum_{m=1}^M\log{p_m}}{n}}.
\eeq
Moreover, if $t\gtrsim (-\log(\kappa))^{-1}\log(n\cdot d_0)$, then 
\bea
\Vert\widehat{\mbB}^{(t)}-\mbB^*\Vert\lesssim \sqrt{\frac{s\sum_{m=1}^M\log{p_m}}{n}}.
\eea
\end{theorem}

Theorem~\ref{thm.B} implies that, under suitable conditions, DEEM produces an accurate estimate for $\mbB^*$ even in ultra-high dimensions after a sufficiently large number of iterations. The condition that $s=o(\sqrt{n/\sum_m\log{p_m}})$ implies that the model should be reasonably sparse. {Also note that, this rate is derived under Condition~(C1). But so far we are only able to guarantee Condition~(C1) when $s=o[\{n/(\sum_m\log{p_m})\}^{1/12}]$ (c.f~Lemma~\ref{lem.C1}), which necessarily implies that $s=o(\sqrt{n/\sum_m\log{p_m}})$. }

We further require $\Delta_0$ to be sufficiently large such that all the models of interest have large $\Delta^*$. To avoid excessively lengthy expressions and calculations, we do not calculate the explicit dependence of our upper bound on $\Delta^*$ here. But we give an intuitive explanation on the impact of $\Delta^*$. Note that \eqref{eq.B.t} contains two terms, $\kappa^t d_0$ and $\sqrt{\frac{s\sum_{m=1}^M\log{p_m}}{n}}$, where $d_0$ is the distance between the initial value and the true parameters. Since $0<\kappa<1/2$, $\kappa^t d_0$ vanishes as long as $t\rightarrow\infty$, but $\Delta^*$ is related to how fast this convergence is. Loosely speaking, the value of $\Delta^*$ inversely affects $\kappa$. For a larger $\Delta^*$, we can find a smaller $\kappa$ such that \eqref{eq.B.t} holds with a high probability, and thus $\widehat\mbB^{(t)}$ converges to $\mbB^*$ in fewer iterations. When $\Delta^*$ is small, we can only find a larger $\kappa$, and the algorithmic convergence is slower. In our theory, $\Delta_0$ can be viewed as the lower bound for $\Delta^*$ such we can find a $\kappa<1/2$ to guarantee \eqref{eq.B.t} with a high probability. See Section~\ref{Sec: Delta} for a numerical demonstration of the effect of $\Delta^*$.

Now we present our main results concerning the clustering error.
 Denote the clustering error of DEEM as
\beq
R(\text{DEEM})=\min_{\Pi: \{1,2\}\mapsto \{1,2\}}\Pr\left(\Pi(\widehat{Y}_i^{\text{DEEM}})\ne Y_i\right).
\eeq
Note that the clustering error is defined as the minimum over all permutations $\Pi: \{1,2\}\mapsto \{1,2\}$, since there could be label switching in clustering.
In the meantime, recall that the lowest clustering error possible is achieved by assigning $\mbX_i$ to Cluster 2 if and only if \eqref{eq.opt} is true. Define the error rate of the optimal clustering rule as
\beq
R(\text{Opt})=\Pr(\widehat{Y}_i^{opt}\ne Y_i),
\eeq
where $\widehat{Y}_i^{opt}$ is determined by the optimal rule in \eqref{eq.opt}. We study $R(\text{DEEM})-R(\text{Opt})$.

\begin{theorem}\label{thm.Err}
Under the conditions in Theorem~\ref{thm.B}, we have that
\begin{enumerate}
\item
For the $\kappa$ that satisfies \eqref{eq.B.t}, if $t\gtrsim (-\log(\kappa))^{-1}\log(n\cdot d_0)$, then with a probability greater than $1-O(\prod p_m^{-1})$, we have
\beq\label{thm.Err.eq1}
R(\text{DEEM})-R(\text{Opt})\lesssim \frac{s\sum_{m=1}^M\log{p_m}}{n}.
\eeq
\item The convergence rate in \eqref{thm.Err.eq1} is minimax optimal over $\boltheta\in \bolTheta(c_{\pi},C_b,s,\{C_m\}_{m=1}^M,\Delta_0)$.
\end{enumerate}
\end{theorem}

Theorem~\ref{thm.Err} shows that the error rate of DEEM converges to the optimal error rate even when the dimension of each mode of the tensor, $p_m$, grows at an exponential rate of $n$. Moreover, the convergence rate is minimax optimal. These results provide strong theoretical support for DEEM. The proofs of the upper bounds in Theorems~\ref{thm.B} \& \ref{thm.Err} are related to those in \citet{Cai2017CHIME}, but require a significant amount of additional efforts. We consider the tensor normal distribution, but non-asymptotic bounds for our estimators of $\bolSigma_1^*,\ldots,\bolSigma_M^*$ are not available in the literature. Also, for us to claim the minimax optimality in Theorem~\ref{thm.Err}, we have to find the lower bound for the excessive clustering error.  This is achieved by constructing a family of models that characterize the intrinsic difficulty of estimating TNMMs. We consider models with sparse means and covariance matrices $\bolSigma_m$ proportional to identity matrices. The excessive clustering error of these models is no smaller than $O(n^{-1}s\sum_{m=1}^M\log{p_m})$.  Because this lower bound matches our upper bound in \eqref{thm.Err.eq1}, we obtain the minimax optimality.


\subsection{Cluster separation}\label{Sec: Delta}
Recall that we define the cluster separation as $\Delta^*=\langle\bolmu_2^*-\bolmu_1^*,\llbracket\bolmu_2^*-\bolmu_1^*;(\bolSigma^*)_1^{-1},\ldots,(\bolSigma_M^*)^{-1}\rrbracket\rangle$. It quantifies the difficulty of clustering, and affects how fast the algorithmic error vanishes throughout the iterations (c.f~Theorem~\ref{thm.B}). Here we demonstrate this impact with a numerical example.

We consider M1 from the simulation (Section~\ref{Sec: num}) as a baseline. Define the cluster separation in M1 as $\Delta^*_1$. We examine the performance of DEEM and its competitors with varying $\Delta^* = a\Delta_1^*$, where $a\in\{0.5,0.75,1,2,3,4\}$. To achieve the specified $\Delta^*$, we proportionally rescale $\bolmu_2^*$ by $\sqrt{a}$ while keeping $\pi_k^*,\bolSigma_m^*$ unchanged. Since the sparse K-means (SKM; \cite{Witten2010SKM}) and DEEM are the top two methods under model M1, we plot the clustering error of SKM, DEEM and the optimal rule in Figure~\ref{fig:M1_delta}. Clearly, both DEEM and SKM have smaller clustering error as $\Delta^*$ increases (left panel), and the relative clustering error shrinks at the same time (middle panel). Therefore, $\Delta^*$ is indeed a very accurate measure of the difficulty of a clustering problem. Moreover, the right panel shows that DEEM needs fewer iterations to achieve convergence when $\Delta^*$ is larger, which confirms our discussion following Theorem~\ref{thm.B}.

\begin{figure}[h!]
	\begin{center}
		\begin{minipage}{0.329\textwidth}
			\centering
			\includegraphics[width=1\textwidth]{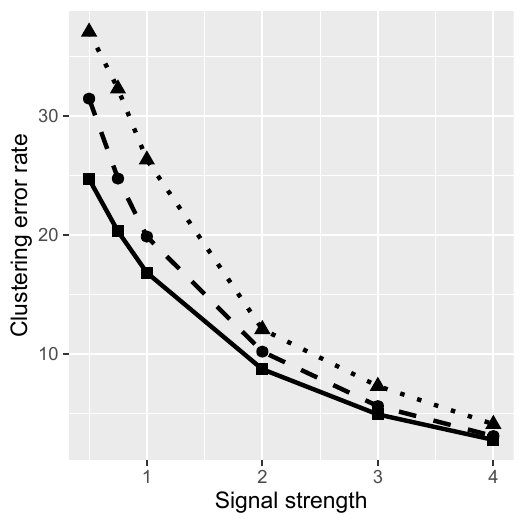}
		\end{minipage}
		\begin{minipage}{0.329\textwidth}
			\centering
			\includegraphics[width=1\textwidth]{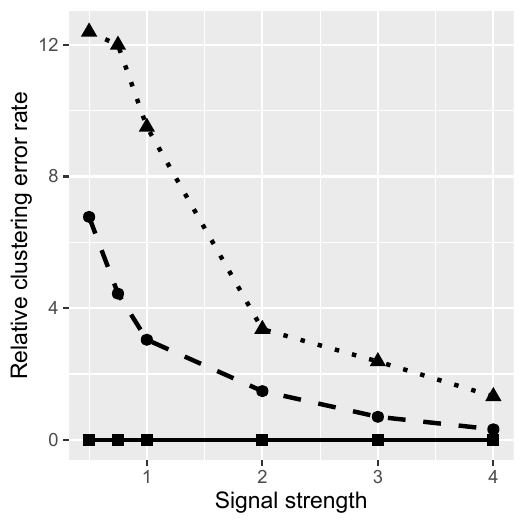}
		\end{minipage}
		\begin{minipage}{0.329\textwidth}
			\centering
			\includegraphics[width=1\textwidth]{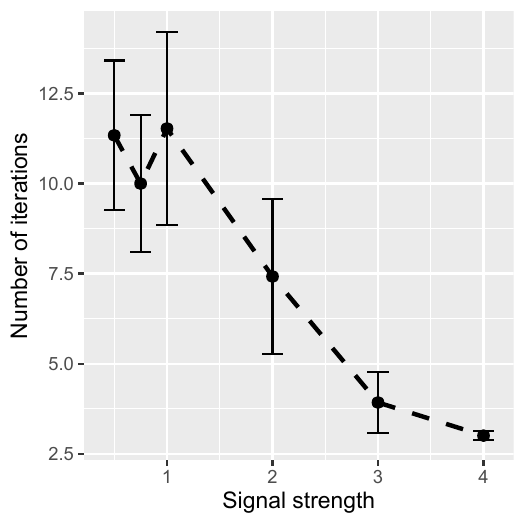}
		\end{minipage}
		\caption{Clustering performance under M1 with varying $\Delta^* = a \times \Delta_1^*$ based on 100 replications. In all panels, the results for SKM are drawn in dotted line, those for DEEM are in dashed line, and those for the optimal rule is in solid line. The left panel shows the clustering error rates $R$ of SKM, DEEM and the optimal rule. The middle panel shows the relative clustering error rates $R-R(\text{Opt})$ of SKM, DEEM and the optimal rule, where $R(\text{Opt})$ is the optimal error rate. The right panel shows number of iterations needed for convergence in DEEM with error bars represent $1.96$ times standard error.}
		\label{fig:M1_delta}
	\end{center}
\end{figure}


\section{Numerical Studies} \label{Sec: num}

\subsection{Simulations}\label{Sec: sim}
In this section, our observations in all models are three-way tensors $\mbX \in \mbbR^{p_1 \times p_2 \times p_3}$. The prior probabilities are set to be $\pi_k^*={1}/{K}$, where $K$ is the number of clusters. For simplicity, we let $n_k$ be equal for $k=1,\ldots,K$ in each model. We fix $\bolmu_1^*=0$, and specify covariance matrices $\bolSigma^*_m$, $m=1,2,3$ and $\mbB_k^*, k=2,\ldots,K$ for each model. For $\mbB_k^*$, all the elements not mentioned in the following model specification are set to be 0. For a matrix $\bolOmega=[\omega_{ij}]$ and a scalar $\rho>0$, we say that $\bolOmega=AR(\rho)$ if $\omega_{ij}=\rho^{|i-j|}$; and we say that $\bolOmega=CS(\rho)$ if $\omega_{ij}=\rho+(1-\rho)\mathrm{1}(i=j)$. 

For each of the following seven simulation settings, we generate 100 independent data sets under the TNMM in $\eqref{margdis}$. Each cluster has sample size $n_k=50$ for Models M5 and M6, and $n_k=75$ for all other models. 
Specifically, the simulation model parameters are as follows.

\noindent{\bf M1}: $K=2$, $p=10\times10\times4$. $\bolSigma_1^*=CS(0.3)$, $\bolSigma_2^*=AR(0.8)$, $\bolSigma_3^*=CS(0.3)$, $\mbB^*_{2,[1:6,1,1]}=0.5$.

\noindent{\bf M2}: Same as {\bf M1} except for $\bolSigma^*_2$, which is specified as follows. 
Let $\bolOmega_0=(\omega_{ij})$ where $\omega_{ij}=u_{ij}\delta_{ij}$, $\delta_{ij}\sim \text{Bernoulli}(1,0.05)$ and $u_{ij}\sim \text{Unif}[0.5,1]\cup [-1,-0.5]$. The we symmetrize $\bolOmega_0$ by setting $\bolOmega=({\bolOmega}_0+{\bolOmega}_0\T)/2$. Set $\bolOmega^*={\bolOmega}+\{\max(-\lambda_{\min}({\Omega}),0)+0.05\}\mbI_{p_2}$. Finally rescale $\bolOmega^*$ such that diagonal elements are 1, and $(\bolSigma_2^*)^{-1}=\bolOmega^*$.

\noindent{\bf M3}: Same as {\bf M1} except for $K=3$, $\bolSigma^*_3=CS(0.5)$ and $\mbB^*_{[2,1:6,1,1]}=-\mbB^*_{[3,1:6,1,1]}=0.5$.

\noindent{\bf M4}:  $K=4,p=10\times 10\times 4$, $\bolSigma^*_1=\mbI_{p_1}$, $\bolSigma^*_2=AR(0.8)$, $\bolSigma^*_3=\mbI_{p_3}$, $\mbB^*_{[2,1:6,1,1]}=-\mbB^*_{[3,1:6,1,1]}=0.8$.

\noindent{\bf M5}: $K=6$, $p=10\times 10\times 4$. $\bolSigma_1^*=AR(0.9)$, $\bolSigma_2^*=CS(0.6)$, $\bolSigma_3^*=AR(0.9)$. $\mbB^*_{[2,1:6,1,1]}=0.6$, $\mbB^*_{[3,1:6,1,1]}=1.2$, $\mbB^*_{[4,1:6,1,1]}=1.8$, $\mbB^*_{[5,1:6,1,1]}=2.4$, $\mbB^*_{[6,1:6,1,1]}=3$.

\noindent{\bf M6}: $K=6$, $p=10\times 10\times 4$.  We specify $\bolmu_k^*$ instead of $\mbB_k^*$. The corner $u_1\times u_2\times u_3 = 8\times1\times1$ sub-tensor of $\bolmu_k$ is filled with independently $\text{Unif}[0,1]$ numbers, while we fill in zeros elsewhere. Then we center it as $\bolmu_k^{*} = \bolmu_k-\bolmu_1$ for $k=1,\dots,K$. 
The covariance matrices $\bolSigma_m^{*}$'s are all two-block-diagonal, where the block sizes corresponding to the zero versus nonzero in means. Each block is generated as $\mbO\mbD\mbO^T$, where $\mbO$ is a randomly generated orthogonal matrix and $\mbD$ is a diagonal matrix that contains the eigenvalues. 
The first block's $\mbD$ is set as $5u$, $u=1,\ldots,u_m$ and the second block's $\mbD$ is set as $2\times\log(v+1)$, $v=1,\dots,p_m-u_m$. Finally we standardize $\bolSigma_m^{*}$ to have unit Frobenius norm. 


\noindent{\bf M7}: $K=2$, $p=30\times30\times30$. $\bolSigma_1^*=CS(0.5)$, $\bolSigma_2^*=AR(0.8)$, $\bolSigma_3^*=CS(0.5)$. $\mbB^*_{[2,1:6,1,1]}=0.6$.



Models M1--M7 cover a wide range of models. In M1 and M2, we consider two mixtures, where we include various covariance structure such as auto-correlation $AR(\rho)$, compound symmetric $CS(\rho)$, and sparse inverse covariance (in M2).
Then in M3 and M4, we increase the number of clusters to $K=3$ and slightly modify other parameters to keep the optimal clustering error around $0.2$.
In M5 and M6, we further increase the number of clusters to $K=6$ and decrease the cluster size $n_k$ from $75$ to $50$. 
In M6, we consider a type of mean-covariance joint parameterization that  corresponds to the envelope mixture models \citep{wang2020model}. This mimics strong correlation but separable signals.
Finally, M7 is constructed so that $p=30^3=27,000$ is significantly higher than the other models.

\begin{table}[t!]
	\centering
		\begin{tabular}[t]{@{}cccccccccc@{}}
			\toprule
			&Optimal &K-means &SKM &{\bf DEEM} &DTC &TBM & {\bf EM}  &AFPF &CHIME\\ 
			\midrule
			\multirow{2}{*}{M1} &16.81 &32.43 &26.31 &19.85 &34.10 &32.91 &34.38 &32.69 &32.45\\
			&(0.34) &(0.40) &(0.68) &(0.35) &(0.42) &(0.38) &(0.42) &(0.39) &(0.41)\\
			\midrule
			\multirow{2}{*}{M2} &9.59 &31.26 &32.01 &12.99 &34.91 &31.43 &28.20 &42.44 &46.75\\
			&(0.25) &(0.42) &(0.67) &(0.53) &(0.87) &(0.41) &(0.54) &(0.66) &(0.24)\\
			\midrule
			\multirow{2}{*}{M3} &17.27 &34.57 &22.32 &20.16 &40.72 &34.75 &32.84 &35.88 &NA\\
			&(0.25) &(0.39) &(0.29) &(0.33) &(0.42) &(0.34) &(0.35) &(0.35) &(--)\\
			\midrule
			\multirow{2}{*}{M4} &22.31 &44.62 &40.21 &26.84 &45.28 &45.78 &42.74 &42.89 &NA\\
			&(0.27) &(0.42) &(0.56) &(0.39) &(0.65) &(0.39) &(0.41) &(0.42) &(--)\\
			\midrule
			\multirow{2}{*}{M5} &8.47 &24.53 &15.93 &10.07 &64.24 &20.78 &19.64 &21.88 &NA\\
			&(0.16) &(0.67) &(0.28) &(0.26) &(0.38) &(0.33) &(0.33) &(0.33) &(--)\\
			\midrule
			\multirow{2}{*}{M6} &10.40 &34.69 &23.93 &16.00 &71.18 &34.13 &27.36 &29.16 &NA\\
			&(0.16) &(0.79) &(0.60) &(0.47) &(0.33) &(0.59) &(0.46) &(1.50) &(--)\\
			\midrule
			\multirow{2}{*}{M7} &8.30 &34.08 &25.85 &12.27 &44.05 &33.61 &33.48 &NA &NA\\
			&(0.20) &(0.64) &(1.17) &(0.74) &(0.51) &(0.63) &(0.64) &(--) &(--)\\
			\bottomrule
		\end{tabular}
		\caption{Reported are the averages and standard errors (in parentheses) of clustering error rates based on 100 replicates. \label{tb:clust_err}}
\end{table}

We consider several popular methods as competitors of DEEM, including K-means, and standard EM (EM; Section~\ref{Sec:EM.issues}), sparse K-means (SKM; \cite{Witten2010SKM}), adaptive pairwise fusion penalized clustering (APFP; \cite{Guo2010}), high-dimensional Gaussian mixtures with EM algorithm (CHIME; \cite{Cai2017CHIME}), dynamic tensor clustering (DTC; \cite{Sun2018}), tensor block model (TBM; \cite{wang2019multiway}). We want to remark that the most direct competitor is the standard EM for TNMM. The DTC and TBM methods are designed for tensor data but from a different perspective. As discussed in the Section~\ref{Sec: Intro}, DTC's advantage is from tensor decomposition and TBM is a co-clustering method (clustering variables and observations simultaneously). Other methods are designed for vector data. We vectorize the tensors before applying the vector-based methods. We use the built-in function in $\tt R$ for K-means, the $\tt R$ package $\tt sparcl$ for SKM, the $\tt R$ package $\tt PARSE$ for APFP, and the $\tt R$ package $\tt tensorsparse$ for TBM. The code of DTC is downloaded from the authors' websites. In addition, we include the error rates of the optimal rule as a baseline.

The implementation of TBM works on three-way data tensor. Our data is a four way tensor of dimension $n\times p_1\times p_2\times p_3$, with the observations being an additional mode. Hence, when we apply TBM, we first apply mode-1 matricization to each observation and then combine the observations as a three-way tensor of dimension $n\times p_1\times (p_2p_3)$. Also, TBM requires specifying the number of clusters along each mode. We use true $K$ as number of clusters along the first mode (i.e, the mode of the observations) and apply the BIC in \citet{wang2019multiway} to tune the numbers of clusters on the second and the third mode. 
In Section~D of Supplementary Materials, we also conduct additional simulations under the TBM data generating process. 

We compare the clustering error rates of all the methods. 
We calculate the clustering error rate to be $\min_{\Pi}\frac{1}{n}\sum_{i=1}^n \mathrm{1}(\widehat Y_i \ne \Pi(Y_i) )$ over all possible permutations $\Pi: \{1,\ldots,K\}\mapsto \{1,\ldots,K\}$ of cluster labels. The clustering error rates are summarized in Table~\ref{tb:clust_err}. Due to its excessively long computation time, the results of AFPF  are based on 30 replications in M6, and are not reported for M7. The results for CHIME are only reported for M1 and M2, because M3--M6 have $K>2$ clusters, while the implementation for CHIME is only available for $K=2$. On the other hand, CHIME exceeds the memory limit of 16GB we set for all methods for M7. 

We make a few remarks on Table~\ref{tb:clust_err}. First of all, DEEM is significantly better than all the other methods across a wide range of TNMM parameter settings. Such results suggest that DEEM has very competitive numerical performance in the presence of different correlation structure, number of clusters and dimensions. The advantage of DEEM is likely a consequence of exploiting the tensor structure, modeling the correlation and imposing variable selection, as no competitor combines all these three components together. 
Secondly, the tensor methods DTC and TBM assume different statistical models and do not account for  the correlation among variables. Therefore, they are less efficient than DEEM under the TNMM. 
Finally, variable selection generally improves clustering accuracy in high dimensions. DEEM and EM fit the same model, but a major distinction between them is that DEEM enforces variable selection while EM does not. Analogously, the sparse K-means (SKM) is uniformly better than K-means. This demonstrates the importance of variable selection in clustering problems.

\subsection{Real data illustration} \label{Sec: real}

We further compare DEEM with the competitors on the BHL (brain, heart and lung) dataset, available at \url{https://www.ncbi.nlm.nih.gov/sites/GDSbrowser?acc=GDS1083}. This dataset contains the expression levels of 1124 genes on 27 brain, heart or lung tissues. On each tissue, the measurement is repeated four times. Hence, our observation $\mbX_i\in\mathbb{R}^{4\times 1124}$, with each row being the gene expression level of one measurement. We attempt to recover the type of each tissue based on $\mbX_i,i=1,\ldots,27$.

We preprocess the data by performing the  Kolmogorov-Smirnov test (KS test) on each column to compare its overall distribution with the normal distribution. Only the columns with small $p$-values are preserved for clustering. We consider reducing the dimension of $\mbX_i$ to $4\times 20$ and $4 \times 30$. We apply DEEM along with all the competitors in Section~\ref{Sec: sim} on this dataset. For DEEM, we generate 30 different initial values and use BIC to tune the initial value along with the tuning parameter. The same is done for DTC. For APFP, it is suggested by the authors to first fit  GMM for 100 times without penalty with different random initial values and select the one with the highest likelihood. We follow this suggestion. The implementations of SKM do not allow users to specify initial values, so we let it pick its own initial value. The clustering error rates are reported in Table~\ref{tab: gds1083}. It can be seen that DEEM has comparable or superior performance to all the competitors in both dimensions. The lowest clustering error rate is achieved by DEEM with dimension $4\times 20$.

\begin{table}[t!]
	\centering
		\begin{tabular}[t]{@{}ccccccccc@{}}
			\toprule
			&DEEM &K-means &SKM &DTC &TBM & EM &AFPF\\ 
			\midrule
			$4\times 20$ &\textbf{7.41} &14.81 &14.81 &22.22 &33.33 &14.81 &14.81 \\
			$4\times 30$ &11.11 &11.11 &11.11 &18.52 &11.11 &11.11 &11.11 \\
			\bottomrule
		\end{tabular}
			\caption{Clustering error rates of the BHL data. \label{tab: gds1083}}
\end{table}

\section{Discussion} \label{Sec: discuss}

In this paper, we propose and study the tensor normal mixture model (TNMM). It is a natural extension of the popular GMM to tensor data. The proposed method simultaneously performs variable selection, covariance estimation and clustering for tensor mixture models. While Kronecker tensor covariance structure is utilized to significantly reduce the number of parameters, it incorporates the dependence between variables and along each tensor modes. This distinguishes our method from independence clustering methods such as K-means. 
We enforce variable selection in the enhanced E-step via convex optimization, where sparsity is directly  derived from the optimal clustering rule.
We propose completely explicit updates in the enhanced M-step, where the new moment-based estimator for covariance is computationally fast and does not require sparsity or other structural assumptions on the covariance. Encouraging theoretical results are established for DEEM, and are further supported by numerical examples. 

Our DEEM algorithm is developed for multi-cluster problem, e.g.~$K\geq 2$, and has been shown to work well in simulations when $K$ is not too large. Since the number of parameters in TNMM grows with $K$, extensions such as low-rank decomposition on $\mbB^*_k$ may be needed for problems where the number of clusters are expected to be large.  Moreover, theoretical study is challenging for $K>2$ and for unknown $K$. Such extensions of our theoretical results  from $K=2$ to general $K$ are yet to be studied. Relatedly, consistent selection of $K$ remains an open question for TNMM. 


\baselineskip=1pt
\bibliographystyle{agsm}
\bibliography{draft}
\end{document}